\documentclass[12pt]{article}
\usepackage[pdfpagemode=UseNone,pdfstartview={XYZ null null null}]{hyperref}
\usepackage[longnamesfirst,numbers,square]{natbib}
\usepackage[margin=0.5in]{geometry} % Smaller margin
\usepackage{ulem}
\usepackage{graphics}
\usepackage{graphicx}
\usepackage{tabularx}
\usepackage{amssymb}
\usepackage{amsmath}
\usepackage{latexsym}
\usepackage{amstext}
\usepackage{mathtools}
\usepackage{amsfonts}
\usepackage{bm}
\usepackage{setspace}
\usepackage{float}
\usepackage{caption}
\usepackage{subcaption}
\usepackage{verbatim}
\usepackage{color}
\newcommand{\p}{\partial}

\newcommand{\beq}{\begin{eqnarray}}
\newcommand{\beqq}{\begin{eqnarray*}}
\newcommand{\eeq}{\end{eqnarray}}
\newcommand{\eeqq}{\end{eqnarray*}}
\newcommand{\eps}{\varepsilon}

\newcommand{\bxi}{\mbox{\boldmath$\xi$}}
\newcommand{\bta}{\mbox{\boldmath$\eta$}}
\newcommand{\x}{\mbox{\boldmath$x$}}

\newcommand{\y}{\mbox{\boldmath$y$}}

\newcommand{\n}{\mbox{\boldmath$n$}}

\newcommand\norm[1]{\left\lVert#1\right\rVert}

\newcommand{\CC}{\mathcal{C}}
\newcommand{\HH}{\mathcal{H}}
\newcommand{\GG}{\mathcal{G}}
\renewcommand{\AA}{\mathcal{A}}
\newcommand{\BB}{\mathcal{B}}
\renewcommand{\SS}{\mathcal{S}}
\newcommand{\PP}{\mathcal{P}}

\newcommand{\RR}{\mathcal{R}}
\newcommand{\KK}{\mathcal{K}}
\newcommand{\FF}{\mathcal{F}}
\newcommand{\TT}{\mathcal{T}}
\newcommand{\VV}{\mathcal{V}}

\newcommand{\tx}{\tilde{\mbox{\boldmath$x$}}}
\newcommand{\ty}{\tilde{\mbox{\boldmath$y$}}}
\newcommand{\ggreen}{\tilde{G}_s(\tx;\ty)}
\newcommand{\abs}[1]{|#1|}
\newcommand{\ggreeneval}{\tilde{G}_s(\tx_2;\tx_1)}

\newcommand{\tit}{\tilde{t}}
\newcommand{\tc}{\tilde{c}}
\newcommand{\tci}{\tilde{c}_i}
\newcommand{\tco}{\tilde{c}_1}

\newcommand{\tji}{\tilde{j}_i}
\newcommand{\CI}{\CC_i}
\newcommand{\CO}{\CC_1}

\newcommand{\CII}{\CI^{(1)}}

\newcommand{\COI}{\CO^{(1)}}

\newcommand{\GGREEN}{\GG_s(\bxi;\boldsymbol{\eta})}
\newcommand{\GGREENEVAL}{\GG_s(\bxi_1;\bxi_2)}

\newcommand{\VI}{\VV^{(1)}}

\newcommand{\tv}{\tilde{v}}

\newcommand{\toai}{\tilde{\Omega}_{a_1}}
\newcommand{\toao}{\tilde{\Omega}_{a_2}}

\newcommand{\oxio}{\Omega_{\bxi_2}}
\newcommand{\oxir}{\Omega_{\bxi_r}}
\newcommand{\eparam}{\frac{J}{\alpha}}
\newcommand{\tOmega}{\tilde{\Omega}}
\newcommand{\atOmega}{\left|\tilde{\Omega}\right|}
\newcommand{\Omxi}{\Omega_{\bxi}}
\newcommand{\volOmxi}{\left|\Omega_{\bxi}\right|}
\definecolor{red}{rgb}{1,0,0}
\newtheorem{pres}{Principal Result}

\numberwithin{equation}{section}
\begin{document}
%%%%%%%%%%%%%%%%%%%%%%%%%%%%%%%%%%%%%%%%%%%%%%%%%%%%%%%%%%%%%%%%%%%%%%%%%%%%%%%%%
%%%%%%%%%%%%%%%%%%%%%%%%%%%%%%%%%%%%%%%%%%%%%%%%%%%%%%%%%%%%%%%%%%%%%%%%%%%%%%%%%
\title{Voltage laws in nanodomains revealed by asymptotics and simulations of electro-diffusion equations}
\author{F. Paquin-Lefebvre$^1$, A. Barea Moreno$^1$ and D. Holcman$^{1,\,2}$ \footnote{$^{1}$ Group of Data Modeling and Computational Biology, IBENS, Ecole Normale Sup\'erieure, 75005 Paris, France. $^2$ Department of Applied Mathematics and Theoretical Physics (DAMTP) and Churchill College, University of Cambridge,  CB30DS, United Kingdom.}}
\date{\today}
\maketitle
%%%%%%%%%%%%%%%%%%%%%%%%%%%%%%%%%%%%%%%%%%%%%%%%%%%%%%%%%%%%%%%%%%%%%%%%%%%%%%%%%
%%%%%%%%%%%%%%%%%%%%%%%%%%%%%%%%%%%%%%%%%%%%%%%%%%%%%%%%%%%%%%%%%%%%%%%%%%%%%%%%%
\begin{abstract}
Characterizing the local voltage distribution within nanophysiological domains, driven by ionic currents through membrane channels, is crucial for studying cellular activity in modern biophysics, yet it presents significant experimental and theoretical challenges. Theoretically, the complexity arises from the difficulty of solving electro-diffusion equations in three-dimensional domains. Currently, there are no methods available for obtaining asymptotic computations or approximated solutions of nonlinear equations, and numerically, it is challenging to explore solutions across both small and large spatial scales. In this work, we develop a method to solve the Poisson-Nernst-Planck equations with ionic currents entering and exiting through two narrow, circular window channels located on the boundary. The inflow through the first window is composed of a single cation, while the outflow maintains a constant ionic density satisfying local electro-neutrality conditions. Employing regular expansions and Green's function representations, we derive the ionic profiles and voltage drops in both small and large charge regimes. We explore how local surface curvature and window channels size influence voltage dynamics and validate our theoretical predictions through numerical simulations, assessing the accuracy of our asymptotic computations. These novel relationships between current, voltage, concentrations and geometry can enhance the characterization of physiological behaviors of nanodomains.
\end{abstract}
%%%%%%%%%%%%%%%%%%%%%%%%%%%%%%%%%%%%%%%%%%%%%%%%%%%%%%%%%%%%%%%%%%%
%%%%%%%%%%%%%%%%%%%%%%%%%%%%%%%%%%%%%%%%%%%%%%%%%%%%%%%%%%%%%%%%%%%
\section{Introduction}
%%%%%%%%%%%%%%%%%%%%%%%%%%%%%%%%%%%%%%%%%%%%%%%%%%%%%%%%%%%%%%%%%%%
Studying voltage in nanodomains has been a continuous effort, leading to the modeling and refinement of transistors to microchips in the field of electronic \cite{bazant}, but has also been key to better characterize the role of subcellular domains in cell biology \cite{koch1989}. Indeed, the voltage in such small domains controls many fundamental physiological and metabolic processes, such as the opening and closing of ionic protein channels \cite{bezanilla2008,hille}, cellular homeostasis \cite{Turrigiano} or ATP production in mitochondria \cite{emboJ,voet,garcia}. However at such nanometer scale, intracellular voltage measurement remains difficult, despite recent experimental advances using nanopipettes \cite{Jayant,Lagacheyuste,JeffHugh2023} or genetically encoded voltage indicators \cite{cohen2012,cartaillerNeuron,Dieudonne}. \\
Classical approaches to model neuronal excitability and voltage propagation rely on a Hodgkin-Huxley formalism \cite{huxley1952} and the cable or telegraphic equations \cite{rall1962}, which treat the plasma membrane as a capacitor with ionic channels as electrical conductors and assume spatially homogeneous ionic concentrations. However these theories do not account for the complex geometry of neuronal nano-structures, such as synapses and dendritic spines, where channel fluxes can lead to ionic concentration gradients and localized spatial perturbations \cite{rusakov2017,holcman2015nrn}. These changes are small and negligible when the volume is large. But for nanodomains with volumes of the order of femtoliters ($\mu{\rm m}^3$), we can calculate that in a volume of one $\mu{\rm m}^3$, an inward current of 1 pA corresponds to an ionic influx of around 10 $\mu{\rm M}$ per millisecond. Thus, currents of the order of tens or hundreds of pA can significantly alter ionic concentrations within milliseconds. \\
To address how subcellular geometry and concentration changes contribute to voltage dynamics at such a small scale, alternative modeling approaches have been developed. One example is electro-diffusion theory describing the motion of charged particles within electrolytes \cite{rubinstein,qian1989}. Mathematically it is formulated with the Poisson coupled to Fokker-Planck equations, called Poisson-Nernst-Planck equations: the voltage is computed from the local charge interactions using Poisson, while spatio-temporal dynamics of ionic densities follow Fokker-Planck (drift-diffusion) equations with the electric field driving the drift. These equations have been widely studied in one-dimensional space, with ionic concentrations following Boltzmann distribution allowing a model reduction to a single nonlinear Poisson's equation. This is known as Poisson--Boltzmann theory describing voltage distribution and local charge imbalances near planar membranes \cite{andelman,BenYaakov2009,orland2000}. However this geometry and steady-state simplification cannot be used to model ionic fluxes originating from channels and pumps in complex three-dimensional subcellular nanodomains, which is the topic of the current study. \\
Previous works either focused on the single-charge non-electroneutral case \cite{cartailler2017jns,cartailler2017physD,cartailler2019jmb,cartailler2017scirep} or two-monovalent charges \cite{cartailler2018neuron,tricot2021,paquin2024}, and highlighted for instance how highly curved membrane protrusions, such as a funnel-shaped cusp, or how the membrane organization of channels, can modulate voltage dynamics. However realistic nanophysiological domains often involve a more complex interplay of multiple ionic species with varying valences. For instance voltage nanodomains can result of the coactivation of calcium and potassium channels \cite{zamponi2011} (Ca$_v$--K$_v$), with calcium influx forcing the exit of potassium by creating a voltage nanodomain. The set of valences involved here are $\left\{+2,+1,-1\right\}$, the negative valence referring to chloride ions. Our aim here is to study how $n$ ionic species of valence $z_i$ with $i=1,\ldots,n$ interact and generate voltage nanodomains between entry and exit channels.\\
The manuscript is organized as follows. We first provide a brief summary of the main result expressed as current-voltage formulas (Section \S \ref{sec:summary}). We formulate the Poisson-Nernst-Planck electro-diffusion model in Section \S \ref{sec:model}. We employ a regular asymptotic expansion to derive ionic and voltage solutions in Section \S \ref{sec:asym}. These computations rely on a Green's function representation of solutions. We find that ionic concentrations at the influx location deviate from local electro-neutrality (\S \ref{sec:small_alpha}), but that these deviations become negligible as the Debye length shrinks (\S \ref{sec:big_alpha}). In Section \S \ref{sec:numerics} we provide numerical simulations on spheroid domains, deforming spheres into prolate or oblate spheroids to study how voltage dynamics is modulated by the local membrane curvature, for the case of two and three ionic charges. Finally, Section \S \ref{sec:perspec} discusses the present computation in the context of neuroscience, with a few open mathematical problems mentioned.
%%%%%%%%%%%%%%%%%%%%%%%%%%%%%%%%%%%%%%%%%%%%%%%%%%%%%%%%%%%%%%%%%%%
\section{Main result summary}\label{sec:summary}
%%%%%%%%%%%%%%%%%%%%%%%%%%%%%%%%%%%%%%%%%%%%%%%%%%%%%%%%%%%%%%%%%%%
The main result of this article is summarized as follows: we obtain relations between ionic densities, voltage, and an influx current composed of single cation ($z_1 > 0$). The bounded domain $\Omega$ (Fig.~\ref{fig:fig0}) is of characteristic length-scale
\beq
R = \frac{1}{2}\max\left.\left\{\|\x - \y\| \right| \x,\,\y \in \p\Omega \right\}\,,
\eeq
with an isoperimetric ratio satisfying $|\p\Omega|/|\Omega|^{2/3} \sim O(1)$. The boundary $\p\Omega$ contains two narrow disk channels $\p\Omega_{A_1}$, $\p\Omega_{A_2}$ of radii $A_1,\,A_2 \ll R$, centered in $\x_1,\,\x_2$, which receive and emit ionic fluxes. The mean membrane curvature in $\x \in \p\Omega$ is given by $H(\x)$ and $\alpha$ is the ratio
\beq
\alpha=\frac{R}{\lambda_D},
\eeq
between R and the Debye length $\lambda_D$. With the above domain characteristics, we use the electro-diffusion model to obtain the following results:
%%%%%%%%%%%%%%%%%%%%%%%%%%%%%%%%%%%%%%%%%%%%%%%%%%%%%%%%%%%%%%%%%%%
\begin{figure}[!ht]
\centering
\includegraphics[width=0.66\linewidth]{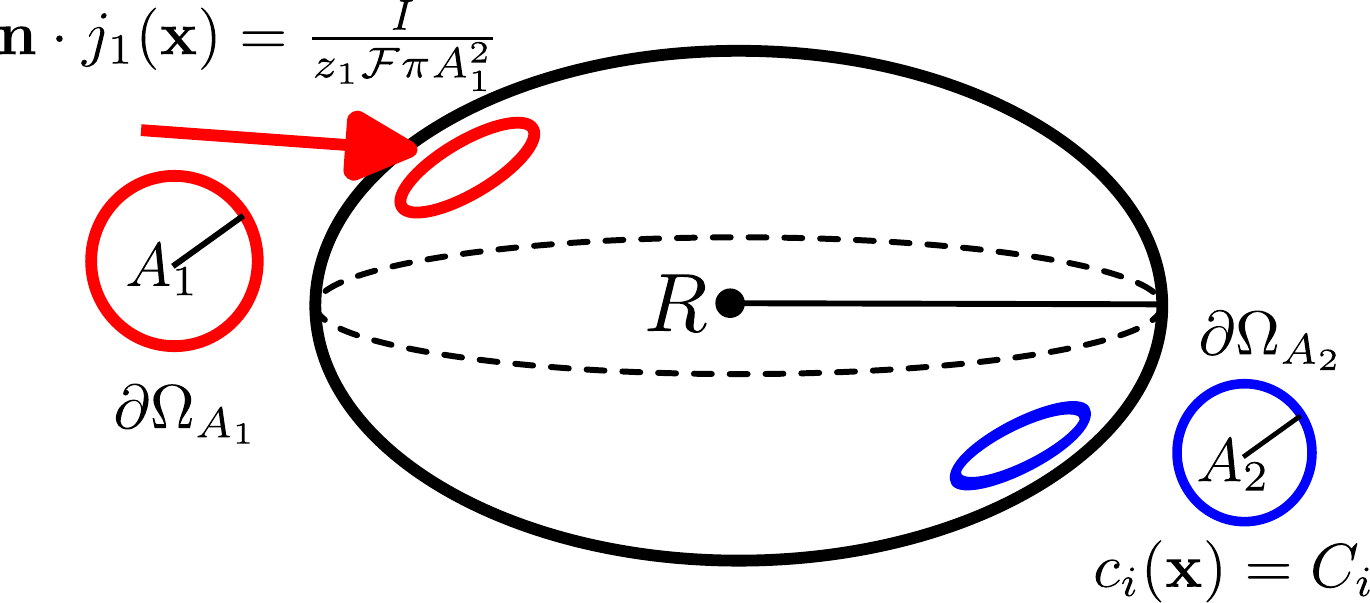}
\caption{\label{fig:fig0} \textbf{Domain schematic diagram.} Example of an ellipsoid-shaped domain $\Omega$ of length-scale $R$. A positive-charge influx current is sent through a first narrow window $\p\Omega_{A_1}$, while on $\p\Omega_{A_2}$ ionic concentrations are set to be constant.}
\end{figure}
%%%%%%%%%%%%%%%%%%%%%%%%%%%%%%%%%%%%%%%%%%%%%%%%%%%%%%%%%%%%%%%%%%%
%%%%%%%%%%%%%%%%%%%%%%%%%%%%%%%%%%%%%%%%%%%%%%%%%%%%%%%%%%%%%%%%%%%
\paragraph{Small charge regime: $\alpha \sim O(1)$.} The ionic and voltage solutions are given at the influx location by
\begin{subequations}\label{eq:alpha_small}
\begin{align}
c_1(\x_1) =& \, C_1 + \frac{I}{z_1\FF D_1 \pi A_2}\left(\frac{A_2}{A_1} - \frac{H(\x_1)}{4}A_2\log\left(\frac{A_1}{R}\right) \right. \nonumber \\
& \left. + \left(1 - \frac{z_1^2C_1}{\sum_{k=1}z_k^2C_k}\right)\left(\frac{\pi}{4} - \frac{H(\x_2)}{4}A_2\log\left(\frac{A_2}{R}\right)\right) + O\left( \frac{A_2}{R} \right) \right) \,, \label{eq:alpha_small_c1} \\
c_i(\x_1) =& \, C_i - z_iC_i\left(\frac{I}{\left(\sum_{k=1}^nz_k^2C_k\right) \FF D_1 \pi A_2 } \left( \frac{\pi}{4} - \frac{H(\x_2)}{4}A_2\log\left(\frac{A_2}{R}\right) + O\left(\frac{A_2}{R}\right)\right)\right), \quad i \neq 1 \,, \label{eq:alpha_small_ci} \\
v(\x_1) =& \, \frac{k_B\TT}{e}\left(\frac{I}{\left(\sum_{k=1}^nz_k^2C_k\right) \FF D_1\pi A_2 }\left( \frac{\pi}{4} - \frac{H(\x_2)}{4}A_2\log\left(\frac{A_2}{R}\right) + O\left(\frac{A_2}{R}\right)\right)\right)\,, \label{eq:alpha_small_v}
\end{align}
\end{subequations}
where we have the Faraday constant $\FF$, thermal voltage $\frac{k_B\TT}{e}$, valence $z_i$ and fixed ionic densities $C_i$ on the exit. Note that $H(\x_1)$ and $H(\x_2)$ are the mean boundary curvature at the center of the narrow windows $\p\Omega_{A_1}$, $\p\Omega_{A_2}$ (Fig.~\ref{fig:fig0}) and that we impose the local electro-neutrality condition
\beq
\sum_{i=1}^n z_iC_i = 0.
\eeq
%%%%%%%%%%%%%%%%%%%%%%%%%%%%%%%%%%%%%%%%%%%%%%%%%%%%%%%%%%%%%%%%%%%
\paragraph{Large charge regime: $\alpha \gg 1$.} The solutions Eq.~\eqref{eq:alpha_small} become
\begin{subequations}\label{eq:alpha_large}
\begin{align}
c_1(\x_1) &= C_1 + \frac{I F(A_1,A_2,R)}{z_1 \FF D_1 \pi A_2 }\,, \label{eq:alpha_large_c1} \\
c_i(\x_1) &= C_i - z_iC_i\left(\frac{I F(A_1,A_2,R)}{\left(\sum_{k=1}^nz_k^2C_k\right)\FF D_1 \pi A_2}\right), \quad i \neq 1 \,, \label{eq:alpha_large_ci} \\
v(\x_1) &= \frac{k_B\TT}{e}\left(\frac{IF(A_1,A_2,R)}{\left(\sum_{k = 1}^n z_k^2 C_k\right) \FF D_1 \pi A_2} \right)\,, \label{eq:alpha_large_v}
\end{align}
\end{subequations}
where
\beq
F(A_1, A_2, R) = \frac{A_2}{A_1} + \frac{\pi}{4} - \frac{A_2}{4} \left(H(\x_1)\log\left(\frac{A_1}{R}\right) + H(\x_2)\log\left(\frac{A_2}{R}\right)\right) + O\left(\frac{A_2}{R}\right)\,.
\eeq
For two-charge models with monovalent ionic species, or more generally if $z_2 = -1$, the voltage is given by
\beq
v(\x_1) = \frac{k_B\TT}{e}\log\left(1 + \frac{IF(A_1,A_2,R)}{\left(\sum_{k = 1}^n z_k^2 C_k\right) \FF D_1 \pi A_2} \right)\,.
\eeq
%%%%%%%%%%%%%%%%%%%%%%%%%%%%%%%%%%%%%%%%%%%%%%%%%%%%%%%%%%%%%%%%%%%
%%%%%%%%%%%%%%%%%%%%%%%%%%%%%%%%%%%%%%%%%%%%%%%%%%%%%%%%%%%%%%%%%%%
\section{PNP electro-diffusion model with multiple ionic charges}\label{sec:model}
%%%%%%%%%%%%%%%%%%%%%%%%%%%%%%%%%%%%%%%%%%%%%%%%%%%%%%%%%%%%%%%%%%%
We describe here the Poisson-Nernst-Planck (PNP) model which consists of the Poisson's equation for the voltage $v(\x,t)$ generated by the local ionic concentration differences, and the Fokker-Planck equations describing the motion of charged particles within the induced electric field. The ionic system within the bounded domain $\Omega$ involves multiple species, each of density $c_i(\x,t)$ and valence $z_i$ for $i = 1,\ldots, n$, with the corresponding electro-chemical gradient flux
\beq
j_i(\x,\,t) = - D_i \left(\nabla c_i(\x,\,t) + \frac{z_i e}{k_B \TT} c_i(\x,\,t) \nabla v(\x,\,t)\right), \quad i = 1,\, \ldots,\, n\,,
\eeq
where $D_i$ is the diffusion coefficient while $k_B\TT/e \approx 26$ mV is the thermal voltage. The Nernst-Planck equation for each $i = 1,\, \ldots,\, n$ is
\beq\label{eq:NP}
\frac{\partial c_i(\x,\,t)}{\partial t} = -\nabla \cdot j_i(\x,\,t) = D_i\left(\Delta c_i(\x,t) + \frac{z_ie}{k_B\TT} \nabla(c_i(\x,t)\nabla v(\x,t)) \right), \quad \x \in \Omega, \quad t > 0\,,
\eeq
while the voltage $v(\x,t)$ is solution of the Poisson's equation
\beq\label{eq:Poiss}
\Delta v(\x,\,t) + \frac{\FF}{\eps\eps_0} \sum_{i = 1}^n z_i c_i(\x,\,t) = 0, \quad \x \in \Omega, \quad t > 0,
\eeq
with Faraday constant $\FF$ and electrical permittivity $\eps\eps_0$.
%%%%%%%%%%%%%%%%%%%%%%%%%%%%%%%%%%%%%%%%%%
\subsection{Boundary conditions}
We model the inward and outward ionic fluxes through membrane-bound protein channels, that are approximated as narrow disks on the boundary $\p\Omega_{A_1}$ and $\p\Omega_{A_2}$, of radius $A_1$, $A_2$ and centered in $\x_1$, $\x_2$. The first window receives an ionic influx of species $c_1(\x,t)$ only, whose valence $z_1$ is positive to model the injection of cations within a neuronal nano-domain, while on the exit site we impose constant ionic densities. Thus on $\p\Omega_{A_1}$ we have the following (Neumann) flux boundary condition,
\begin{subequations}\label{eq:pnp_BC}
\begin{align}
\n \cdot j_1(\x,t) + \frac{I}{z_1 \FF \pi A_1^2} = 0, \quad \n \cdot j_i(\x,t) = 0, \quad \x \in \p\Omega_{A_1}, \quad i=2,\ldots,n\,, \label{eq:influx}
\end{align}
where $\n$ is the outward unit vector to $\p\Omega$. On $\p\Omega_{A_2}$ we set each ionic densities to be constant, with the rest of the boundary $\p\Omega_r = \p\Omega\backslash\{\p\Omega_{A_1}\cap\p\Omega_{A_2}\}$ that is fully reflecting to all ions, thus yielding
\begin{align}
c_i(\x,t) = C_i, \quad \x \in \p\Omega_{A_2}, \quad \n \cdot j_i(\x,t) = 0, \quad \x \in \p\Omega_r, \quad i = 1,\ldots,n\,, \label{eq:cte_ref}
\end{align}
to which we add the local electro-neutrality condition,
\begin{align}
\sum_{i=1}^n z_iC_i = 0\,.
\end{align}
Lastly the boundary conditions for the voltage is imposed on the exit site $\p\Omega_{A_2}$ as ground condition, while everywhere else we neglect the capacitance of the membrane and set to zero the component of the electric field normal to the boundary, leading to
\begin{align}
\x \in \p\Omega_{A_2}, \quad \n \cdot \nabla v(\x,t) = 0\,, \quad \x \in \p\Omega\backslash\p\Omega_{A_2}\,, \quad v(\x,t) = 0\,.
\end{align}
\end{subequations}
%%%%%%%%%%%%%%%%%%%%%%%%%%%%%%%%%%%%%%%%%%%%%%%%%%%%%%%%%%%%%%%%%%%
\subsection{Non-dimensionalization}
%%%%%%%%%%%%%%%%%%%%%%%%%%%%%%%%%%%%%%%%%%%%%%%%%%%%%%%%%%%%%%%%%%%
We show here how we use the dimensional spatial $\tx$ and time $\tit$ variables:
\begin{equation}
\tx = \frac{\x}{R} \in \tilde{\Omega},\, \quad \tit = \frac{D_1}{R^2} t
\end{equation}
where $D_1$ is the diffusion coefficient of the first ionic species, while $\tilde{\Omega} = \Omega / R$ is the rescaled domain. The weighted sum $\sum_{k=1}^n z_k^2C_k$ measures the conductivity of the electrolyte. By using this term and the thermal voltage $k_B\TT/e$ we can define new density and voltage variables given by
\begin{equation}
\tci(\tx,\,\tit) = \frac{c_i\left(R \tx,\, \frac{R^2}{D_1} \tit\right)}{\sum_{k=1}^n z_k^2 C_k},\, \quad i = 1,\, \ldots,\, n,\, \quad \tv(\tx,\,\tit) = \frac{e}{k_B\TT}v\left(R \tx,\, \frac{R^2}{D_1} \tit\right)\,,
\end{equation}
as well as dimensionless electro-chemical gradients,
\begin{equation}
\tji(\tx,\,\tit) = \frac{R j_i\left(R \tx,\, \frac{R^2}{D_1} \tit\right)}{D_i\sum_{k=1}^n z_k^2 C_k} = -\left(\nabla c_i(\tx,\tit) + z_i c_i(\tx,\tit)\nabla \tv(\tx,\tit)\right),\, \quad i = 1,\, \ldots,\, n\,.
\end{equation}
Then, upon substituting within the PNP Eq.~\eqref{eq:NP}-\eqref{eq:pnp_BC}, we obtain
\begin{subequations}\label{eq:pnp_tilde}
\begin{align}
\tau_i\frac{\partial \tci(\tx,\,\tit)}{\partial t} + \nabla \cdot \tji(\tx,\,\tit) = 0,\, \quad i = 1,\, \ldots,\, n, \quad &\tx \in \tilde{\Omega}\,, \\
\Delta \tv(\tx,\,\tit) + \alpha^2 \sum_{i = 1}^n z_i \tci(\tx,\,\tit) = 0,\, \quad &\tx \in \tilde{\Omega}\,,
\end{align}
\end{subequations}
where $\tau_i = D_1/D_i$ is a diffusion coefficients ratio while $\alpha$ is given by
\beq
\alpha = \frac{R}{\lambda_D}\,, \quad \text{with} \quad \lambda_D = \sqrt{\frac{k_B \TT \eps \eps_0}{e \FF \sum_{k=1}^n z_k^2C_k}}\,,
\eeq
and thus it corresponds to the ratio of the domain length-scale $R$ over the Debye length $\lambda_D$ measuring electro-static interaction lengths in the absence of external flux. The boundary conditions Eq.~\eqref{eq:pnp_BC} get reformulated as
\begin{subequations}\label{eq:pnp_BC_tilde}
\begin{align}
\n \cdot \tilde{j}_1(\tx,\,\tit) + \frac{J}{z_1} = 0,\, \quad \n \cdot \tci(\tx,\,\tit) = 0,\, \quad i = 2,\, \ldots,\, n,\, \quad \n \cdot \tv(\tx,\,\tit) = 0,\, \quad &\tx \in \partial \toai,\\
\tci(\tx,\,\tit) = \nu_i,\, \quad i = 1,\, \ldots,\, n,\, \quad \tv(\tx,\,\tit) = 0,\, \quad &\tx \in \partial \toao, \\
\n \cdot \tilde{j}_i(\tx,\,\tit) = 0,\, \quad i = 1,\, \ldots,\, n,\, \quad \n \cdot \tv(\tx,\,\tit) = 0,\, \quad &\tx \in \partial \tilde{\Omega}_r,
\end{align}
\end{subequations}
where $\partial \toai$ and $\partial \toao$ are two small circular windows of radii $a_1 = A_1/R$ and $a_2=A_2/R$, centered in $\tx_1 = \x_1/R$ and $\tx_2 = \x_2/R$. In Eq.~\eqref{eq:pnp_BC_tilde} the parameter $J$ measures the strength of the influx current entering window $\partial \toai$, while $\nu_i$ for each $i$ is proportional to the exit site $\partial \toao$ constant ionic density:
\begin{equation}
J = \frac{I R}{\FF \pi A_1^2 D_1 \left(\sum_{k=1}^n z_k^2C_k\right)},\, \quad \nu_i = \frac{C_i}{\sum_{k=1}^n z_k^2C_k},\, \quad i = 1,\,\ldots,\,n \,.
\end{equation}
We seek solutions for the ionic densities $\tci(\tx)$ and voltage $\tv(\tx)$ distribution at steady-state, thus satisfying
\begin{subequations}\label{eq:ss_eqs}
\begin{align}
\nabla \cdot \tji(\tx) = \Delta \tci(\tx) + z_i \nabla \cdot (\tci(\tx) \nabla \tv(\tx)) = 0,\, \quad &\tx \in \tilde{\Omega}, \\ \Delta \tv(\tx) + \alpha^2 \sum_{i = 1}^n z_i \tci(\tx) = 0,\, \quad &\tx \in \tilde{\Omega}\,,
\end{align}
\end{subequations}
for two different parameter regimes: first $\alpha \sim O(1)$, and then $\alpha \gg 1$. For the first case the Debye length is of the order of the domain length-scale, while for the latter $\lambda_D$ is much smaller than $R$. Length-scales of neuronal nano-compartments are usually of the order of hundreds of nanometers (with $R=500$ nm for a typical dendritic spine head), while for a 1:1 binary electrolyte with steady-state ionic density $C_1=C_2=100$ mM, we calculate $\lambda_D \approx 1$ nm. Thus in practice $\alpha \gg 1$, but we will show that our analysis can also treat the case $\alpha \sim O(1)$ corresponding to small interior charge densities. The two distinct cases can be handled by introducing the coordinate $\bxi = \alpha \tx$ within the stretched domain $\Omega_{\bxi} = \alpha \tilde{\Omega}$, but we remark that this is the same as rescaling $\x\in\Omega$ by the Debye length, with $\bxi = \x / \lambda_D$. Upon defining new density and voltage variables as
\begin{equation}
\CI(\bxi) = \tci(\bxi/\alpha),\, \quad i = 1,\, \ldots,\, n,\, \quad \VV(\bxi) = \tv(\bxi/\alpha)\,.
\end{equation}
we can reformulate \eqref{eq:ss_eqs} as
\begin{subequations}\label{eq:pnp_stretched}
\begin{align}
\Delta \CI(\bxi) + z_i \nabla (\CI(\bxi) \nabla \VV(\bxi)) = 0,\, \quad &\bxi \in \Omega_{\bxi}\,, \\
\Delta \VV(\bxi) + \sum_{i = 1}^n z_i \CI(\bxi) = 0,\, \quad &\bxi \in \Omega_{\bxi}\,,
\end{align}
subject to the boundary conditions
\begin{align}
\frac{\partial \CO(\bxi)}{\partial \n} = \frac{J}{\alpha z_1}, \quad \frac{\partial \CI(\bxi)}{\partial \n} = 0, \quad i = 2,\, \ldots,\, n,\, \quad \frac{\partial \VV(\bxi)}{\partial \n} = 0, \quad &\bxi \in \partial \Omega_{\bxi_1}\,, \\
\CI(\bxi) = \nu_i,\, \quad i = 1,\, \ldots,\, n,\, \quad \VV(\bxi) = 0,\, \quad &\bxi \in \partial \Omega_{\bxi_2}\,, \\
\frac{\partial \CI(\bxi)}{\partial \n} = 0, \quad i = 1,\, \ldots,\, n,\, \quad \frac{\partial \VV(\bxi)}{\partial \n} = 0, \quad &\bxi \in \partial \Omega_{\bxi_r}\,,
\end{align}
\end{subequations}
where $\p\Omega_{\bxi_j}$ for $j=1,\,2$ are two circular boundary disks of radius $\alpha a_j$ centered in $\bxi_j$ on $\p\Omega_{\bxi}$. Note that the ratio $\eparam$ is the single parameter controlling the strength of the influx and the electric potential due to ionic density disturbances.
%%%%%%%%%%%%%%%%%%%%%%%%%%%%%%%%%%%%%%%%%%%%%%%%%%%%%%%%%%%%%%%%%%%
\section{Introducing a regular asymptotic expansion}\label{sec:asym}
%%%%%%%%%%%%%%%%%%%%%%%%%%%%%%%%%%%%%%%%%%%%%%%%%%%%%%%%%%%%%%%%%%%
To derive an approximate solution to system Eq.~\eqref{eq:pnp_stretched}, we introduce a regular asymptotic expansion in terms of the ratio $\eparam$ around the background steady-state and locally electro-neutral ionic solutions,
\begin{subequations}\label{eq:expan}
\begin{align}
\CI(\bxi) &= \nu_i + \CII(\bxi) \eparam + O\left(\left(\frac{J}{\alpha}\right)^2 \right), \quad i = 1,\, \ldots,\, n\,, \\
\VV(\bxi) &= \VI(\bxi) \eparam + O\left(\left(\frac{J}{\alpha}\right)^2 \right)\,,
\end{align}
\end{subequations}
which is valid assuming $J \ll \alpha$. Thus when $\alpha \sim O(1)$ with the Debye length of the order of the domain length-scale, we must have $J \ll 1$. Alternatively if $\alpha \gg 1$ the expansion \eqref{eq:expan} will also hold for a current $J$ of order one, with $J \sim O(1)$. By collecting terms at order $O\left(\eparam\right)$ we get the following linearized problem,
\begin{subequations}\label{eq:order1}
\begin{align}
\Delta \CII(\bxi) + z_i \nu_i \Delta \VI(\bxi) &= 0, \quad \bxi \in \Omega_{\bxi}\,, \\
\Delta \VI(\bxi) + \sum_{i = 1}^n z_i \CII(\bxi) &= 0, \quad \bxi \in \Omega_{\bxi}\,, \end{align}
along with the boundary conditions
\begin{align}
\frac{\partial \COI(\bxi)}{\partial \n} = \frac{1}{z_1}, \quad \frac{\partial \CII(\bxi)}{\partial \n} = 0, \quad i = 2,\, \ldots,\, n,\, \quad \frac{\partial \VI(\bxi)}{\partial \n} = 0, \quad &\bxi \in \partial \Omega_{\bxi_1}\,, \\
\CII(\bxi) = 0, \quad i = 1,\, \ldots,\, n, \quad \VI(\bxi) = 0,\, \quad &\bxi \in \partial \oxio\,, \\
\frac{\partial \CII(\bxi)}{\partial \n} = 0, \quad i = 1,\, \ldots,\, n, \quad \frac{\partial \VI(\bxi)}{\partial \n} = 0, \quad &\bxi \in \partial \Omega_{\bxi_r}\,.
\end{align}
\end{subequations}
We solve this system by introducing the intermediate functions below
\begin{equation}
\SS_{ij}(\bxi) = z_j \nu_j \CII(\bxi) - z_i \nu_i \CC_{j}^{(1)}(\bxi), \quad \PP(\bxi) = \sum_{i = 1}^n z_i \CII(\bxi),
\end{equation}
where we note that $\SS_{ij}(\bxi) = -\SS_{ji}(\bxi)$ and $\SS_{ii}(\bxi) = 0$, and with $\SS_{ij}(\bxi)$ and $\PP(\bxi)$ satisfying
\begin{subequations}\label{eq:pde_ss_pp}
\begin{align}
\Delta \SS_{ij}(\bxi) = 0, \quad \Delta \PP(\bxi) - \PP(\bxi) = 0\,.
\end{align}
Next, boundary conditions then are established by considering the different indices $i,\,j$,
\begin{align}
\frac{\partial \SS_{1j}(\bxi)}{\partial \n} &= \frac{z_j\nu_j}{z_1},\, \quad \bxi \in \partial \Omega_{\bxi_1},\, \quad \SS_{1j}(\bxi) = 0,\, \quad \bxi \in \partial \oxio,\, \quad \frac{\partial \SS_{1j}(\bxi)}{\partial \n} = 0,\, \quad \bxi \in\partial \oxir, \quad j \neq 1, \\
\frac{\partial \SS_{j1}(\bxi)}{\partial \n} &= - \frac{z_j\nu_j}{z_1},\, \quad \bxi \in \partial \Omega_{\bxi_1},\, \quad \SS_{j1}(\bxi) = 0,\, \quad \bxi \in \partial \oxio,\, \quad \frac{\partial \SS_{j1}(\bxi)}{\partial \n} = 0,\, \quad \bxi \in \partial \oxir, \quad j \neq 1, \\
\frac{\partial \SS_{ij}(\bxi)}{\partial \n} &= 0,\, \quad \bxi \in \partial \Omega_{\bxi_1},\, \quad \SS_{ij}(\bxi) = 0,\, \quad \bxi \in \partial \oxio,\, \quad \frac{\partial \SS_{ij}(\bxi)}{\partial \n} = 0,\, \quad \bxi \in \partial \oxir, \quad i \neq 1,\, j \neq 1,
\end{align}
from which we easily establish that $\SS_{ij}(\bxi) = 0$ when both $i \neq 1$ and $j \neq 1$. We then find that the function $\PP(\bxi)$ has the boundary condition
\begin{align}
\frac{\partial \PP(\bxi)}{\partial \n} = 1,\, \quad \bxi \in \partial \Omega_{\bxi_1},\, \quad \PP(\bxi) = 0,\, \quad \bxi \in \partial \oxio,\, \quad \frac{\partial \PP(\bxi)}{\partial \n} = 0,\, \quad \bxi \in \partial \oxir\,.
\end{align}
\end{subequations}
Although fluxes through the exit $\partial \Omega_{\bxi_2}$ are unknown, we show in Appendix \S \ref{sec:weber} that they can be approximated by the classical Weber solution \cite{crank1975},
\begin{equation}\label{eq:weber}
\frac{\p\SS_{1j}(\bxi)}{\p\n} = \frac{\KK_{1j}}{\sqrt{(\alpha a_2)^2 - \norm{\bxi - \bxi_2}^2}},\, \quad j \neq 1, \quad \frac{\p\PP(\bxi)}{\p\n} = \frac{\KK}{\sqrt{(\alpha a_2)^2 - \norm{\bxi - \bxi_2}^2}},\, \quad \bxi \in \p\oxio\,,
\end{equation}
where the constants $\KK_{1j}$ and $\KK$ are evaluated upon applying the divergence theorem to Eq.~\eqref{eq:pde_ss_pp} and \eqref{eq:weber},
\begin{align}
\pi (\alpha a_1)^2 \frac{z_j\nu_j}{z_1} + 2 \pi \KK_{1j} \alpha a_2 &= 0 \,, \\
\pi (\alpha a_1)^2 + 2 \pi \KK \alpha a_2 - \abs{\Omega_{\bxi}} \overline{\PP}  &= 0 \,,
\end{align}
yielding
\begin{equation}\label{eq:eq_KK}
\KK_{1j} = -\frac{\alpha a_1^2z_j\nu_j}{2a_2z_1}\,, \quad \KK = \frac{\volOmxi\overline{\PP}}{2\pi\alpha a_2} - \frac{\alpha a_1^2}{2a_2}\,.
\end{equation}
We solve the system \eqref{eq:pde_ss_pp} with the Green's function $\GGREEN$ (Appendix \S \ref{sec:green}) solution of
\begin{equation}\label{eq:green_bxi_main}
\Delta \GGREEN = \frac{1}{\volOmxi}, \quad \bxi \in \Omxi, \quad \frac{\p \GGREEN}{\p \n} = \delta(\bxi - \bta), \quad \bxi \in \p \Omxi, \quad \int_{\Omxi}\GGREEN d\bxi = 0 \quad \bta \in \p \Omega_{\bxi},
\end{equation}
Upon applying Green's second identity to the systems \eqref{eq:pde_ss_pp} and \eqref{eq:green_bxi_main}, we obtain
\begin{subequations}\label{eq:greenid}
\begin{align}
\SS_{1j}(\bta) &= \overline{\SS_{1j}} + \int_{\partial \Omega_{\bxi_1}}\GGREEN \frac{\partial \SS_{1j}(\bxi)}{\partial \n} d\bxi + \int_{\partial \oxio} \GGREEN \frac{\partial \SS_{1j}(\bxi)}{\partial \n} d\bxi\,, \label{eq:greenid_ss} \\
\PP(\bta) &= \overline{\PP} + \int_{\partial \Omega_{\bxi_1}} \GGREEN \frac{\partial \PP(\bxi)}{\partial \n} d\bxi + \int_{\partial \Omega_{\bxi_2}} \GGREEN \frac{\partial \PP(\bxi)}{\partial \n} d\bxi - \int_{\Omega_{\bxi}} \GGREEN \PP(\bxi) d\bxi\,, \label{eq:greenid_pp}
\end{align}
\end{subequations}
where $\overline{\SS_{1j}}$ and $\overline{\PP}$ indicate domain averages,
\beq\label{eq:ave}
\overline{\SS_{1j}} = \frac{1}{\volOmxi}\int_{\Omxi} \SS_{1j}(\bxi) d\bxi, \quad \overline{\PP} = \frac{1}{\volOmxi}\int_{\Omxi} \PP(\bxi) d\bxi\,,
\eeq
and then after substituting $\bta = \bxi_1,\, \bxi_2$ and using the absorbing boundary conditions $\SS_{1j}(\bxi_2) = \PP(\bxi_2) = 0$, we obtain
\begin{subequations}\label{eq:greenid_sub}
\begin{align}
\SS_{1j}(\bxi_1) &= \overline{\SS_{1j}} + \frac{z_j\nu_j}{z_1}\int_{\p \Omega_{\bxi_1}}\GG_s(\bxi;\bxi_1)d\bxi + \KK_{1j}\int_{\p \oxio} \frac{\GG_s(\bxi;\bxi_1)}{\sqrt{(\alpha a_2)^2 - \norm{\bxi - \bxi_2}^2}} d\bxi\,, \label{eq:greenid_sub_ss_1} \\
0 &= \overline{\SS_{1j}} + \frac{z_j\nu_j}{z_1}\int_{\p \Omega_{\bxi_1}}\GG_s(\bxi;\bxi_2) d\bxi + \KK_{1j}\int_{\p \oxio} \frac{\GG_s(\bxi;\bxi_2)}{\sqrt{(\alpha a_2)^2 - \norm{\bxi - \bxi_2}^2}} d\bxi\,, \label{eq:greenid_sub_ss_2} \\
\PP(\bxi_1) &= \overline{\PP} + \int_{\p \Omega_{\bxi_1}} \GG_s(\bxi;\bxi_1) d\bxi + \KK \int_{\p\Omega_{\bxi_2}} \frac{\GG_s(\bxi;\bxi_1)}{\sqrt{(\alpha a_2)^2 - \norm{\bxi - \bxi_2}^2}} d\bxi - \int_{\Omega_{\bxi}} \GG_s(\bxi;\bxi_1) \PP(\bxi) d\bxi\,, \label{eq:greenid_sub_pp_1} \\
0 &= \overline{\PP} + \int_{\p\Omega_{\bxi_1}} \GG_s(\bxi;\bxi_2) d\bxi + \KK \int_{\p\Omega_{\bxi_2}} \frac{\GG_s(\bxi;\bxi_2)}{\sqrt{(\alpha a_2)^2 - \norm{\bxi - \bxi_2}^2}} d\bxi - \int_{\Omega_{\bxi}} \GG_s(\bxi;\bxi_2) \PP(\bxi) d\bxi\,. \label{eq:greenid_sub_pp_2}
\end{align}
\end{subequations}
These integrals involving the Neumann Green's function are evaluated in Appendix \S \ref{sec:integrals}, where we get the linear system of equations below for $\SS_{1j}(\bxi_1),\,\overline{\SS_{1j}},\,\PP(\bxi_1)$ and $\overline{\PP}$,
\begin{subequations}
\begin{align}
\SS_{1j}(\bxi_1) &= \overline{\SS_{1j}} + \frac{z_j\nu_j}{z_1} a_1 \alpha f(a_1) + 2 \pi \KK_{1j} a_2\alpha\GG_s(\bxi_2;\bxi_1)\,, \\
0 &= \overline{\SS_{1j}} + \frac{z_j\nu_j}{z_1} \pi (\alpha a_1)^2 \GGREENEVAL + \KK_{1j} \frac{\pi}{2} g(a_2)\,, \\
\PP(\bxi_1) &= \overline{\PP} + a_1 \alpha f(a_1) + 2 \pi \KK a_2\alpha \GG_s(\bxi_2;\bxi_1)  - \PP(\bxi_1) \frac{1}{2}(\alpha a_1)^2m(a_1)\,, \\
0 &= \overline{\PP} + \pi (\alpha a_1)^2 \GGREENEVAL + \KK \frac{\pi}{2} g(a_2) - \PP(\bxi_1) \frac{2\pi}{3}a_1^3\alpha^3\GG_s(\bxi_1;\bxi_2)\,,
\end{align}
\end{subequations}
and then upon substituting expressions for $\KK_{1j}$ and $\KK$ from Eq.~\eqref{eq:eq_KK} we get
\begin{subequations}\label{eq:sys}
\begin{align}
\SS(\bxi_1) - \overline{\SS_{1j}} &= \frac{z_j\nu_j}{z_1} \alpha a_1^2 \left(\frac{f(a_1)}{a_1}-\pi\alpha\GG_s(\bxi_2;\bxi_1)\right)\,, \\
\overline{\SS_{1j}} &= \frac{z_j\nu_j}{z_1} \alpha a_1^2\left(\frac{\pi g(a_2)}{4a_2}-\pi\alpha\GG_s(\bxi_2;\bxi_1)\right) \,, \\
\left(1+\frac{1}{2}(\alpha a_1)^2m(a_1)\right)\PP(\bxi_1) - \left(1 +\left| \Omega_{\bxi} \right|\GG_s(\bxi_2;\bxi_1) \right)\overline{\PP} &= \alpha a_1^2 \left(\frac{f(a_1)}{a_1}-\pi\alpha\GG_s(\bxi_2;\bxi_1)\right) \,, \\
- \frac{2\pi}{3}a_1^3\alpha^3\GG_s(\bxi_1;\bxi_2)\PP(\bxi_1) + \left(1 + \frac{g(a_2)}{4\alpha a_2}\left|\Omega_{\bxi}\right|\right)\overline{\PP} &= \alpha a_1^2\left(\frac{\pi g(a_2)}{4a_2}-\pi\alpha\GG_s(\bxi_2;\bxi_1)\right) \,.
\end{align}
\end{subequations}
Here $f(a_1)$, $g(a_2)$ and $m(a_1)$ are expansions of the narrow window radii $a_1$ and $a_2$, given by
\begin{subequations}
\begin{align}
f(a_1) &= 1 - \alpha\frac{\HH(\bxi_1)}{4} a_1 \log(a_1) + \alpha a_1\left( \pi\RR_s(\bxi_1;\bxi_1) + \frac{\HH(\bxi_1)}{4}\left(\frac{1}{2} - \log(\alpha)\right)\right) + O(a_1^2)\,, \\
g(a_2) &= 1 - \alpha\frac{\HH(\bxi_2)}{\pi} a_2 \log(a_2) + \alpha a_2 \left(4\RR_s(\bxi_2;\bxi_2) + \frac{\HH(\bxi_2)}{\pi}\left(1-\log(2) - \log(\alpha)\right) \right) + O(a_2^2) \,, \\
m(a_1) &= 1 - \alpha\frac{\HH(\bxi_1)}{3} a_1 \log(a_1) + \alpha a_1\left( \frac{4\pi}{3}\RR_s(\bxi_1;\bxi_1) + \frac{\HH(\bxi_1)}{3}\left(\frac{1}{3} - \log(\alpha)\right)\right) + O\left(a_1^2\right)\,.
\end{align}
\end{subequations}
Upon solving the system Eq.~\eqref{eq:sys} we find that $\SS_{1j}(\bxi_1)$ and $\overline{\SS_{1j}}$ are given by
\begin{align}
\SS_{1j}(\bxi_1) &= \frac{z_j\nu_j}{z_1} a_1^2 \alpha \left(\frac{f(a_1)}{a_1} + \frac{\pi g(a_2)}{4 a_2} - 2 \pi \alpha \GGREENEVAL\right)\,, \\
\overline{\SS}_{1j} &= \frac{z_j\nu_j}{z_1} a_1^2 \alpha \left( \frac{\pi g(a_2)}{4 a_2} - \alpha\pi\GGREENEVAL\right)\,,
\end{align}
while for $\PP(\bxi_1)$ and $\overline{\PP}$ we have
\begin{equation}
\PP(\bxi_1) = \alpha a_1^2 \AA, \quad \overline{\PP} = \alpha a_1^2 \BB\,,
\end{equation}
where $\AA$ and $\BB$ are two fractions
\begin{align}
\AA =
\frac{\left(\frac{f(a_1)}{a_1}-\pi\alpha\GG_s(\bxi_2;\bxi_1)\right)\left(1+\frac{g(a_2)}{4\alpha a_2}\left|\Omega_{\bxi}\right|\right) + \left(\frac{\pi g(a_2)}{4a_2} - \pi\alpha\GG_s(\bxi_2;\bxi_1)\right)\left(1 + \left|\Omega_{\bxi}\right|\GG_s(\bxi_2;\bxi_1)\right)}
{\left(1+\frac{1}{2}(\alpha a_1)^2m(a_1) \right)\left(1+\frac{g(a_2)}{4\alpha a_2}\left|\Omega_{\bxi}\right|\right) - \left(1 + \left|\Omega_{\bxi}\right|\GG_s(\bxi_2;\bxi_1)\right)\frac{2\pi}{3}a_1^3\alpha^3\GG_s(\bxi_1;\bxi_2)} \,, \\
\BB =
\frac{\left(1 + \frac{1}{2}(\alpha a_1)^2m(a_1) \right)\left(\frac{\pi g(a_2)}{4a_2} - \pi\alpha\GG_s(\bxi_2;\bxi_1)\right) + \left(\frac{f(a_1)}{a_1}-\pi\alpha\GG_s(\bxi_2;\bxi_1)\right)\frac{2\pi}{3}a_1^3\alpha^3\GG_s(\bxi_1;\bxi_2)}
{\left(1+\frac{1}{2}(\alpha a_1)^2m(a_1) \right)\left(1+\frac{g(a_2)}{4\alpha a_2}\left|\Omega_{\bxi}\right|\right) - \left(1 + \left|\Omega_{\bxi}\right|\GG_s(\bxi_2;\bxi_1)\right)\frac{2\pi}{3}a_1^3\alpha^3\GG_s(\bxi_1;\bxi_2)} \,.
\end{align}
To recover the perturbation terms at $O\left(\eparam\right)$ we simply need to calculate
\begin{align}
\COI(\bxi) &= z_1\nu_1\PP(\bxi) + \sum_{j = 2}^n z_j \SS_{1j}(\bxi), \\
\CII(\bxi) &= z_i\nu_i\PP(\bxi) - z_1\SS_{1i}(\bxi) \quad i \neq 1\,.
\end{align}
At the influx location this yields
\begin{align}
\COI(\bxi_1) &= z_1\nu_1 a_1^2 \alpha\left(\AA + \left(\frac{1}{z_1^2\nu_1} - 1\right)\left(\frac{f(a_1)}{a_1} + \frac{\pi g(a_2)}{4 a_2} - 2 \pi \alpha \GGREENEVAL\right)\right)\,, \\
\CII(\bxi_1) &= z_i\nu_i a_1^2 \alpha\left(\AA - \left(\frac{f(a_1)}{a_1} + \frac{\pi g(a_2)}{4 a_2} - 2 \pi \alpha \GGREENEVAL\right)\right)\,, \quad i  \neq 1\,,
\end{align}
while for the averages we find
\begin{align}
\overline{\COI} &= z_1\nu_1 a_1^2 \alpha\left(\BB + \left(\frac{1}{z_1^2\nu_1} - 1\right)\left(\frac{\pi g(a_2)}{4 a_2} - \pi \alpha \GGREENEVAL\right)\right)\,, \\
\overline{\CII} &= z_i\nu_i a_1^2 \alpha\left(\BB - \left(\frac{\pi g(a_2)}{4 a_2} - \pi \alpha \GGREENEVAL\right)\right)\,, \quad i  \neq 1\,.
\end{align}
Finally we have the first-order approximations for ionic densities at the influx location
\begin{subequations}\label{eq:cinflux}
\begin{align}
\CO(\bxi_1) &= \nu_1 + Jz_1\nu_1 a_1^2 \left(\AA + \left(\frac{1}{z_1^2\nu_1} - 1\right)\left(\frac{f(a_1)}{a_1} + \frac{\pi g(a_2)}{4 a_2} - 2 \pi \alpha \GGREENEVAL\right)\right) + O\left( \left(\frac{J}{\alpha}\right)^2 \right)\,, \\
\CI(\bxi_1) &= \nu_i + Jz_i\nu_i a_1^2 \left(\AA - \left(\frac{f(a_1)}{a_1} + \frac{\pi g(a_2)}{4 a_2} - 2 \pi \alpha \GGREENEVAL\right)\right) + O\left( \left(\frac{J}{\alpha}\right)^2 \right)\,, \quad i  \neq 1\,,
\end{align}
\end{subequations}
as well as averages given by
\begin{subequations}\label{eq:cav}
\begin{align}
\overline{\CO} &= \nu_1 + Jz_1\nu_1 a_1^2 \left(\BB + \left(\frac{1}{z_1^2\nu_1} - 1\right)\left(\frac{\pi g(a_2)}{4 a_2} - \pi \alpha \GGREENEVAL\right)\right) + O\left( \left(\frac{J}{\alpha}\right)^2 \right)\,, \\
\overline{\CI} &= \nu_i + Jz_i\nu_i a_1^2 \left(\BB - \left(\frac{\pi g(a_2)}{4 a_2} - \pi \alpha \GGREENEVAL\right)\right) + O\left( \left(\frac{J}{\alpha}\right)^2\right)\,, \quad i  \neq 1\,.
\end{align}
\end{subequations}
%%%%%%%%%%%%%%%%%%%%%%%%%%%%%%%%%%%%%%%%%%%%%%%%%%%%%%%%%%%%%%%%%%%
\subsection{Asymptotic solution for the voltage}\label{sec:voltage}
%%%%%%%%%%%%%%%%%%%%%%%%%%%%%%%%%%%%%%%%%%%%%%%%%%%%%%%%%%%%%%%%%%%
We proceed by finding a first-order voltage solution. At $O\left(\eparam\right)$ we have
\beq\label{eq:order1V}
\Delta \VI(\bxi) = - \PP(\bxi), \quad \bxi \in \Omxi, \quad \VI(\bxi) = 0, \quad \bxi \in \p\oxio, \quad \frac{\p\VI(\bxi)}{\p\n} = 0, \quad \p\Omxi \backslash \p\oxio\,,
\eeq
with Green's identity that yields
\beq\label{eq:greenV}
\VI(\bta) = \overline{\VI} + \int_{\p\oxio} \GGREEN \frac{\p \VV(\bxi)}{\p\n} d\bxi + \int_{\Omxi} \GGREEN \PP(\bxi) d\bxi\,,
\eeq
where $\overline{\VI}$ corresponds to the domain average as in Eq.~\eqref{eq:ave}. Here also we can use Weber's solution to approximate the flux through $\p\Omxi$,
\beq\label{eq:consV}
\frac{\p\VI(\bxi)}{\p\n} = \frac{\KK}{\sqrt{(\alpha a_2)^2 - \norm{\bxi - \bxi_2}^2}}, \quad \bxi \in \p\oxio, \quad \text{with} \quad \KK = -\volOmxi\frac{\overline{\PP}}{2\pi\alpha a_2}\,,
\eeq
where the constraint on $\KK$ is obtained by applying the divergence theorem to Eq.~\eqref{eq:order1V}. By then setting $\bta = \bxi_1,\bxi_2$ within Eq.~\eqref{eq:greenV} we obtain
\begin{subequations}
\begin{align}
\VI(\bxi_1) &= \overline{\VI} + 2\pi a_2 \KK \alpha \GG_s(\bxi_2;\bxi_1) + \frac{1}{2} (\alpha a_1)^2 m(a_1)\PP(\bxi_1)\,, \\
0 &= \overline{\VI} + \KK\frac{\pi}{2}g(a_2) + \frac{2\pi}{3}a_1^3\alpha^3 \GGREENEVAL\PP(\bxi_1)\,,
\end{align}
\end{subequations}
which becomes,
\begin{subequations}
\begin{align}
\VI(\bxi_1) - \overline{\VI} &= \alpha a_1^2\left(\frac{1}{2} (\alpha a_1)^2 m(a_1)\AA - \volOmxi\GG_s(\bxi_2;\bxi_1)\BB \right)\,, \\
\overline{\VI} &= \alpha a_1^2\left(\frac{g(a_2)}{4\alpha a_2} \volOmxi\BB  - \frac{2\pi}{3}a_1^3\alpha^3\GGREENEVAL \AA \right)\,,
\end{align}
\end{subequations}
after using Eq.~\eqref{eq:consV} with $\PP(\bxi_1) = \alpha a_1^2 \AA$ and $\overline{\PP} = \alpha a_1^2 \BB$. The solution to this system is given by
\begin{subequations}
\begin{align}
\VI(\bxi_1) &= \alpha a_1^2 \left( \left( m(a_1) - \frac{4\pi}{3}a_1\alpha\GGREENEVAL\right)\frac{1}{2}(\alpha a_1)^2\AA + \left(\frac{g(a_2)}{4 a_2} - \alpha\GG_s(\bxi_2;\bxi_1) \right) \frac{\volOmxi}{\alpha}\BB \right)\,, \\
\overline{\VI} &= \alpha a_1^2 \left(\volOmxi\frac{g(a_2)}{4\alpha a_2}\BB - \frac{2\pi}{3}a_1^3\alpha^3\GGREENEVAL\AA\right)\,,
\end{align}
\end{subequations}
and then the first-order approximation becomes
\begin{subequations}\label{eq:VV_gen}
\begin{align}
\VV(\bxi_1) &= Ja_1^2 \left( \left( m(a_1) - \frac{4\pi}{3}a_1\alpha\GGREENEVAL\right)\frac{1}{2}(\alpha a_1)^2\AA + \left(\frac{g(a_2)}{4 a_2} - \alpha\GG_s(\bxi_2;\bxi_1) \right) \frac{\volOmxi}{\alpha}\BB \right) + O\left( \left(\frac{J}{\alpha}\right)^2 \right)\,, \label{eq:VV_gen_influx} \\
\overline{\VV} &= Ja_1^2 \left(\volOmxi\frac{g(a_2)}{4\alpha a_2}\BB - \frac{2\pi}{3}a_1^3\alpha^3\GGREENEVAL\AA\right) + O\left( \left(\frac{J}{\alpha}\right)^2 \right)\,. \label{eq:VV_gen_bar}
\end{align}
\end{subequations}
%%%%%%%%%%%%%%%%%%%%%%%%%%%%%%%%%%%%%%%%%%%%%%%%%%%%%%%%%%%%%%%%%%%
\subsection{Small interior charge parameter regime}\label{sec:small_alpha}
%%%%%%%%%%%%%%%%%%%%%%%%%%%%%%%%%%%%%%%%%%%%%%%%%%%%%%%%%%%%%%%%%%%
We now derive refined asymptotic approximations for the charge densities and voltage valid when $\alpha \sim O(1)$, with the Debye length that is of the same order as the domain characteristic length-scale. Thus for the expansion parameter to be small we must have a small current $J$, and then combined with the narrow window limit this yields the following parameter regime,
\beq
\alpha \sim O(1), \quad J \ll 1, \quad a_1 \ll 1, \quad a_2 \ll 1\,.
\eeq
Upon expanding the ratios $\AA$ and $\BB$ in terms of the narrow window radii and neglecting terms of order $O(1)$ and beyond, we find in Appendix \S \ref{sec:small_alpha_app} that
\begin{equation}
\AA \approx \frac{f(a_1)}{a_1} + \frac{\pi \alpha}{\volOmxi} \approx \frac{1}{a_1} - \alpha\frac{\HH(\bxi_1)}{4} \log(a_1) + O(1), \quad \BB \approx \frac{\pi \alpha}{\volOmxi} \sim O(1)\,,
\end{equation}
and thus the ionic charge densities in Eq.~\eqref{eq:cinflux} become
\begin{align}
\CO(\bxi_1) &\approx \nu_1 + \frac{Ja_1^2}{z_1}\left( \frac{1}{a_1} - \alpha\frac{\HH(\bxi_1)}{4} \log(a_1) + (1 - z_1^2\nu_1)\left(\frac{\pi}{4a_2} - \alpha\frac{\HH(\bxi_2)}{4} \log(a_2)\right) + O(1) \right) + O\left( \left(\frac{J}{\alpha}\right)^2 \right) \,, \\
\CI(\bxi_1) &\approx \nu_i - Jz_i\nu_ia_1^2\left(\frac{\pi}{4a_2} - \alpha\frac{\HH(\bxi_2)}{4} \log(a_2) + O(1) \right) + O\left( \left(\frac{J}{\alpha}\right)^2 \right) \,, \quad i \neq 1\,, \label{eq:cinflux_small_J}
\end{align}
while for the average ionic densities we have
\begin{align}
\overline{\CO} &\approx \nu_1 + \frac{Ja_1^2}{z_1}(1 - z_1^2\nu_1)\left(\frac{\pi}{4a_2} - \alpha\frac{\HH(\bxi_2)}{4} \log(a_2) + O(1) \right) + O\left( \left(\frac{J}{\alpha}\right)^2 \right) \,, \\
\overline{\CI} &\approx \nu_i - Jz_i\nu_ia_1^2 \left(\frac{\pi}{4a_2} - \alpha\frac{\HH(\bxi_2)}{4} \log(a_2) + O(1) \right) + O\left( \left(\frac{J}{\alpha}\right)^2 \right) \,, \quad i \neq 1\,.
\end{align}
Also by performing an asymptotic reduction assuming $a_1,\,a_2 \ll 1$ we find in Appendix \S \ref{sec:small_alpha_app} the exact same current-voltage linear relation at the influx location and for the domain average,
\begin{equation}\label{eq:v_small_alpha}
\VV(\bxi_1) = \overline{\VV} \approx Ja_1^2 \left(\frac{\pi}{4a_2} - \alpha\frac{\HH(\bxi_2)}{4} \log(a_2) + O(1) \right) + O\left( \left(\frac{J}{\alpha}\right)^2 \right)\,. \\
\end{equation}
Interestingly this equation can be recovered by using the classical Boltzmann solution
\begin{equation}
\VV(\bxi) = - \frac{1}{z_i} \log\left(\frac{\CI(\bxi)}{\nu_i}\right) = \log\left(\left(\frac{\CI(\bxi)}{\nu_i}\right)^{-\frac{1}{z_i}}\right), \quad \bxi \in \Omxi, \quad i \neq 1\,,
\end{equation}
which is exact except for the first ion that is injected into the domain. We then set $\bxi = \bxi_1$ and substitute the expansion in Eq.~\eqref{eq:cinflux_small_J} to get
\begin{align*}
\VV(\bxi_1) &= \log\left( \left( 1 - Jz_ia_1^2\left(\frac{\pi}{4a_2} - \alpha\frac{\HH(\bxi_2)}{4} \log(a_2) + O(1) \right) + O\left( \left(\frac{J}{\alpha}\right)^2 \right)  \right)^{-1/z_i}\right), \quad i \neq 1\,,
\end{align*}
but then since $Ja_1^2 \ll 1$ we can use the binomial expansion to remove the dependence on the valence $z_i$
\begin{equation}
\VV(\bxi_1) \approx \log\left(1 + Ja_1^2\left(\frac{\pi}{4a_2} - \alpha\frac{\HH(\bxi_2)}{4} \log(a_2) + O(1) \right) + O\left( \left(\frac{J}{\alpha}\right)^2 \right) \right)\,,
\end{equation}
while with one last Taylor expansion step we recover Eq.~\eqref{eq:v_small_alpha}. We conclude this section with a summary of ionic and voltage formulas on the domain $\tOmega$.
%%%%%%%%%%%%%%%%%%%%%%%%%%%%%%%%%%%%%%%%%%%%%%%%%%%%%%%%%%%%%%%%%%%
\begin{pres}\label{pres:small_alpha}
We consider a steady-state Poisson-Nernst-Planck electro-diffusion model with multiple ionic species $\tc_i(\tx)$ of valence $z_i$ in a bounded domain $\tOmega$, with order one volume $\atOmega \sim O(1)$ and mean boundary curvature function $\tilde{H}(\tx)$ with $\tx \in \p\tOmega$. Assume that a single charge $\tc_1(\x)$ of positive valence $z_1 > 0$ is injected within the domain through a first narrow circular window $\p\tOmega_{a_1}$ while it exits via a second window $\p\tOmega_{a_2}$. Each window has radius $a_j \ll 1$ and is centered in $\tx_j \, \in \p\tOmega$, with $l = \|\tx_1 - \tx_2\| \sim O(1)$ as these windows are well-spaced. In the parameter regime $\alpha \sim O(1)$ with small influx current $J \ll 1$, then the regular asymptotic expansions below hold for ionic densities at the influx location,
\begin{subequations}
\begin{align}
\tco(\tx_1) &= \nu_1 + \frac{Ja_1^2}{z_1}\left(\frac{1}{a_1} - \frac{\tilde{H}(\tx_1)}{4}\log(a_1) + \left(1 - z_1^2\nu_1\right)\left(\frac{\pi}{4a_2} - \frac{\tilde{H}(\tx_2)}{4}\log(a_2)\right) + O(1) \right) + O\left(\left(\frac{J}{\alpha}\right)^2\right) \,, \\
\tci(\tx_1) &= \nu_i - Jz_i\nu_i a_1^2 \left( \frac{\pi}{4a_2} - \frac{\tilde{H}(\tx_2)}{4}\log(a_2) + O(1)\right) + O\left(\left(\frac{J}{\alpha}\right)^2\right), \quad i  \neq 1 \,,
\end{align}
\end{subequations}
while expansions for spatial average densities $\overline{\tc_i}$ are given by
\begin{subequations}
\begin{align}
\overline{\tc_1} &= \nu_1 + \frac{Ja_1^2}{z_1}\left(1 - z_1^2\nu_1\right)\left( \frac{\pi}{4a_2} - \frac{\tilde{H}(\tx_2)}{4}\log(a_2) + O(1) \right) + O\left(\left(\frac{J}{\alpha}\right)^2\right) \,, \\
\overline{\tc_i} &= \nu_i - Jz_i\nu_i a_1^2 \left( \frac{\pi}{4a_2} - \frac{\tilde{H}(\tx_2)}{4}\log(a_2) + O(1)\right) + O\left(\left(\frac{J}{\alpha}\right)^2\right), \quad i  \neq 1 \,.
\end{align}
\end{subequations}
We further calculate that ionic concentrations at the influx location deviate from local electro-neutrality while global deviations are negligible, yielding
\begin{equation}
\sum_{j=1}^n z_j \tc_j(\tx_1) \approx Ja_1^2\left(\frac{1}{a_1} - \frac{\tilde{H}(\tx_1)}{4} \log(a_1)\right) + O\left( Ja_1^2\right), \quad \sum_{j=1}^n z_j \overline{\tc_j} \sim O\left(Ja_1^2\right)\,.
\end{equation}
For the special case of a 1:1 electrolyte with $z_1 = +1$ and $z_2 = -1$ these expansions reduce to
\begin{subequations}
\begin{align}
\tco(\tx_1) &= \frac{1}{2} + \frac{Ja_1^2}{2}\left(\frac{2}{a_1} + \frac{\pi}{4a_2} - \frac{\tilde{H}(\tx_1)}{2}\log(a_1) - \frac{\tilde{H}(\tx_2)}{4}\log(a_2) + O(1) \right) + O\left(\left(\frac{J}{\alpha}\right)^2\right) \,, \\
\tc_2(\tx_1) &= \frac{1}{2} + \frac{Ja_1^2}{2} \left( \frac{\pi}{4a_2} - \frac{\tilde{H}(\tx_2)}{4}\log(a_2) + O(1)\right) + O\left(\left(\frac{J}{\alpha}\right)^2\right) \,,
\end{align}
\end{subequations}
while for average ionic densities we get the exact same expansion,
\begin{equation}
\overline{\tc_1} = \overline{\tc_2} =  \frac{1}{2} + \frac{Ja_1^2}{2}\left( \frac{\pi}{4a_2} - \frac{\tilde{H}(\tx_2)}{4}\log(a_2) + O(1) \right) + O\left(\left(\frac{J}{\alpha}\right)^2\right)\,.
\end{equation}
Finally in this parameter regime voltage and influx current are expected to vary linearly as for Ohm's law, with the voltage drop across the domain and its spatial averages having the same magnitude as revealed by the formula below,
\begin{equation}
\tv(\tx_1) = \overline{\tv} =  Ja_1^2 \left( \frac{\pi}{4a_2} - \frac{\tilde{H}(\tx_2)}{4}\log(a_2) + O(1)\right) + O\left(\left(\frac{J}{\alpha}\right)^2\right) \,.
\end{equation}
\end{pres}
%%%%%%%%%%%%%%%%%%%%%%%%%%%%%%%%%%%%%%%%%%%%%%%%%%%%%%%%%%%%%%%%%%%
\subsection{Large interior charge parameter regime}\label{sec:big_alpha}
%%%%%%%%%%%%%%%%%%%%%%%%%%%%%%%%%%%%%%%%%%%%%%%%%%%%%%%%%%%%%%%%%%%
Alternatively in the regime $\alpha \gg 1$ we can neglect the contribution from the ratios $\AA$ and $\BB$ since we have (see Appendix \ref{sec:big_alpha_app}),
\beq
\AA \sim O\left(\frac{1}{\alpha^2}\right), \quad \BB \sim O\left(\frac{1}{\alpha^2}\right)\,,
\eeq
and then by using the approximation below
\begin{align*}
\frac{f(a_1)}{a_1} + \frac{\pi g(a_2)}{4 a_2} - 2 \pi \alpha \GGREENEVAL \approx \frac{1}{a_1} - \alpha\frac{\HH(\bxi_1)}{4} \log(a_1) + \frac{\pi}{4a_2} - \alpha\frac{\HH(\bxi_2)}{4} \log(a_2) + O(1)
\end{align*}
we find the following formulas for ionic densities at the influx location
\begin{subequations}
\begin{align}
\CO(\bxi_1) &\approx \nu_1 + \frac{Ja_1^2}{z_1}\left(1 - z_1^2\nu_1\right) \left( \frac{1}{a_1} + \frac{\pi}{4a_2} - \alpha\frac{\HH(\bxi_1)}{4} \log(a_1) - \alpha\frac{\HH(\bxi_2)}{4} \log(a_2) + O(1) \right) + O\left(\left(\frac{J}{\alpha}\right)^2\right)\,, \\
\CI(\bxi_1) &\approx \nu_i - Jz_i\nu_i a_1^2 \left( \frac{1}{a_1} + \frac{\pi}{4a_2} - \alpha\frac{\HH(\bxi_1)}{4} \log(a_1) - \alpha\frac{\HH(\bxi_2)}{4} \log(a_2) + O(1)\right) + O\left(\left(\frac{J}{\alpha}\right)^2\right)\,, \quad i  \neq 1 \,, \label{eq:cin_alpha_big}
\end{align}
\end{subequations}
while for the average concentrations we have
\begin{subequations}
\begin{align}
\overline{\CO} &\approx \nu_1 + \frac{Ja_1^2}{z_1}\left(1 - z_1^2\nu_1 \right) \left(\frac{\pi}{4a_2} - \alpha\frac{\HH(\bxi_2)}{4} \log(a_2) + O(1)\right) + O\left(\left(\frac{J}{\alpha}\right)^2\right)\,, \\
\overline{\CI} &\approx \nu_i - Jz_i\nu_i a_1^2 \left(\frac{\pi}{4a_2} - \alpha\frac{\HH(\bxi_2)}{4} \log(a_2) + O(1)\right) + O\left(\left(\frac{J}{\alpha}\right)^2\right)\,, \quad i  \neq 1\,.
\end{align}
\end{subequations}
In Appendix \ref{sec:big_alpha_app} we also evaluate the voltage drop and spatial average to be approximated by
\begin{subequations}
\begin{align}
\VV(\bxi_1) &\approx Ja_1^2 \left(\frac{1}{a_1} + \frac{\pi}{4a_2} - \alpha\frac{\HH(\bxi_1)}{4}\log(a_1) - \alpha\frac{\HH(\bxi_2)}{4}\log(a_2) + O(1) \right) + O\left(\left(\frac{J}{\alpha}\right)^2\right)\,, \label{eq:ohmicV} \\
\overline{\VV} &\approx Ja_1^2 \left( \frac{\pi}{4a_2} - \alpha\frac{\HH(\bxi_2)}{4}\log(a_2) + O(1) \right) + O\left(\left(\frac{J}{\alpha}\right)^2\right)\,. \label{eq:ohmicVbar}
\end{align}
\end{subequations}
As in Section \S \ref{sec:small_alpha}, the classical Boltzmann solution yields
\begin{align*}
\VV(\bxi_1) &= \log\left(\left(\frac{\CI(\bxi_1)}{\nu_i}\right)^{-\frac{1}{z_i}}\right)\,, \quad i \neq 1\,,
\end{align*}
and then by substituting Eq.~\eqref{eq:cin_alpha_big} we get
\begin{align*}\label{eq:v_boltz}
\VV(\bxi_1) =&\, \log\left(\left(1 - Jz_i a_1^2 \left( \frac{1}{a_1} + \frac{\pi}{4a_2} - \alpha\frac{\HH(\bxi_1)}{4} \log(a_1) - \alpha\frac{\HH(\bxi_2)}{4} \log(a_2) + O(1)\right) \right. \right. \\
& \left. \left. \, + O\left(\left(\frac{J}{\alpha}\right)^2\right) \right)^{-\frac{1}{z_i}}\right), \quad i \neq 1\,.
\end{align*}
The dependence upon the valence can be removed by using the binomial expansion,
\begin{align*}
\VV(\bxi_1) &= \log\left(1 + Ja_1^2 \left( \frac{1}{a_1} + \frac{\pi}{4a_2} - \alpha\frac{\HH(\bxi_1)}{4} \log(a_1) - \alpha\frac{\HH(\bxi_2)}{4} \log(a_2) + O(1)\right)+ O\left(\left(\frac{J}{\alpha}\right)^2\right)\right)\,,
\end{align*}
and by expanding the log function we recover formula Eq.~\eqref{eq:ohmicV}. We conclude this section with a summary of ionic and voltage formulas on the domain $\tOmega$ with order one length-scale when $\alpha \gg 1$.
%%%%%%%%%%%%%%%%%%%%%%%%%%%%%%%%%%%%%%%%%%%%%%%%%%%%%%%%%%%%%%%%%%%
\begin{pres}\label{eq:big_alpha}
In the regime $\alpha \gg 1$ with small influx current $J \ll 1$, then the regular asymptotic expansions below hold for ionic densities at the influx location,
\begin{subequations}
\begin{align}
\tco(\tx_1) &= \nu_1 + \frac{Ja_1^2}{z_1}\left(1 - z_1^2\nu_1\right)\left(\frac{1}{a_1} + \frac{\pi}{4a_2} - \frac{\tilde{H}(\tx_1)}{4}\log(a_1) - \frac{\tilde{H}(\tx_2)}{4}\log(a_2) + O(1) \right) + O\left(\left(\frac{J}{\alpha}\right)^2\right) \,, \\
\tci(\tx_1) &= \nu_i - Jz_i\nu_ia_1^2\left(\frac{1}{a_1} + \frac{\pi}{4a_2} - \frac{\tilde{H}(\tx_1)}{4}\log(a_1) - \frac{\tilde{H}(\tx_2)}{4}\log(a_2) + O(1) \right) + O\left(\left(\frac{J}{\alpha}\right)^2\right), \quad i \neq 1 \,,
\end{align}
\end{subequations}
while expansions for spatial averages are given by,
\begin{subequations}
\begin{align}
\overline{\tco} &= \nu_1 + \frac{Ja_1^2}{z_1}\left(1 - z_1^2\nu_1\right)\left( \frac{\pi}{4a_2} - \frac{\tilde{H}(\tx_2)}{4}\log(a_2) + O(1) \right) + O\left(\left(\frac{J}{\alpha}\right)^2\right) \,, \\
\overline{\tci} &= \nu_i - Jz_i\nu_i a_1^2 \left( \frac{\pi}{4a_2} - \frac{\tilde{H}(\tx_2)}{4}\log(a_2) + O(1)\right) + O\left(\left(\frac{J}{\alpha}\right)^2\right), \quad i  \neq 1 \,,
\end{align}
\end{subequations}
which are the same as for the $\alpha \sim O(1)$ regime. Deviations from local or global electro-neutrality are also found to be negligible, as shown upon evaluating the sums,
\beq
\sum_{j=1}^n z_j \tc_j(\tx_1) \sim O\left(Ja_1^2\right), \quad \sum_{j=1}^n z_j \overline{\tc_j} \sim O\left(Ja_1^2\right)\,.
\eeq
Furthermore, the special case of a 1:1 electrolyte yields the expansions below
\begin{subequations}
\begin{align}
\tc_1(\tx_1) &= \frac{1}{2} + \frac{Ja_1^2}{2} \left(\frac{1}{a_1} + \frac{\pi}{4a_2} - \frac{\tilde{H}(\tx_1)}{4}\log(a_1) - \frac{\tilde{H}(\tx_2)}{4}\log(a_2) + O(1) \right) + O\left(\left(\frac{J}{\alpha}\right)^2\right) \,, \\
\tc_2(\tx_1) &= \frac{1}{2} + \frac{Ja_1^2}{2}\left(\frac{1}{a_1} + \frac{\pi}{4a_2} - \frac{\tilde{H}(\tx_1)}{4}\log(a_1) - \frac{\tilde{H}(\tx_2)}{4}\log(a_2) + O(1) \right) + O\left(\left(\frac{J}{\alpha}\right)^2\right)\,,
\end{align}
\end{subequations}
while for ionic spatial averages we get
\begin{equation}
\overline{\tc_1} = \overline{\tc_2} = \frac{1}{2} + \frac{Ja_1^2}{2} \left( \frac{\pi}{4a_2} - \frac{\tilde{H}(\tx_2)}{4}\log(a_2) + O(1) \right) + O\left(\left(\frac{J}{\alpha}\right)^2\right) \,.
\end{equation}
In this regime we find that the total voltage drop across the domain $\tv(\tx_1)$ differs from the average voltage $\overline{\tv}$, as shown by the asymptotic expansions
\begin{subequations}
\begin{align}
\tv(\tx_1) &= Ja_1^2 \left(\frac{1}{a_1} + \frac{\pi}{4a_2} - \frac{\tilde{H}(\tx_1)}{4}\log(a_1) - \frac{\tilde{H}(\tx_2)}{4}\log(a_2) + O(1) \right) + O\left(\left(\frac{J}{\alpha}\right)^2\right) \,, \\
\overline{\tv} &= Ja_1^2 \left( \frac{\pi}{4a_2} - \frac{\tilde{H}(\tx_2)}{4}\log(a_2) + O(1)\right) + O\left(\left(\frac{J}{\alpha}\right)^2\right) \,.
\end{align}
\end{subequations}
which holds for $J \ll 1$. For the special case of a 1:1 electrolyte, or more generally if $z_2 = -1$, then a straightforward evaluation of Eq.~\eqref{eq:v_boltz} yield a logarithmic current-voltage relationship,
\beq
\tv(\tx_1) = \log\left(1 + Ja_1^2 \left(\frac{1}{a_1} + \frac{\pi}{4a_2} - \frac{\tilde{H}(\tx_1)}{4}\log(a_1) - \frac{\tilde{H}(\tx_2)}{4}\log(a_2) + O(1) \right) + O\left(\left(\frac{J}{\alpha}\right)^2\right) \right) \,, \\
\eeq
which also hold in the nonlinear regime.
\end{pres}
%%%%%%%%%%%%%%%%%%%%%%%%%%%%%%%%%%%%%%%%%%%%%%%%%%%%%%%%%%%%%%%%%%%
\section{Numerical simulations on spheroid domains}\label{sec:numerics}
\begin{figure}[!ht]
\centering
\includegraphics[width=\linewidth]{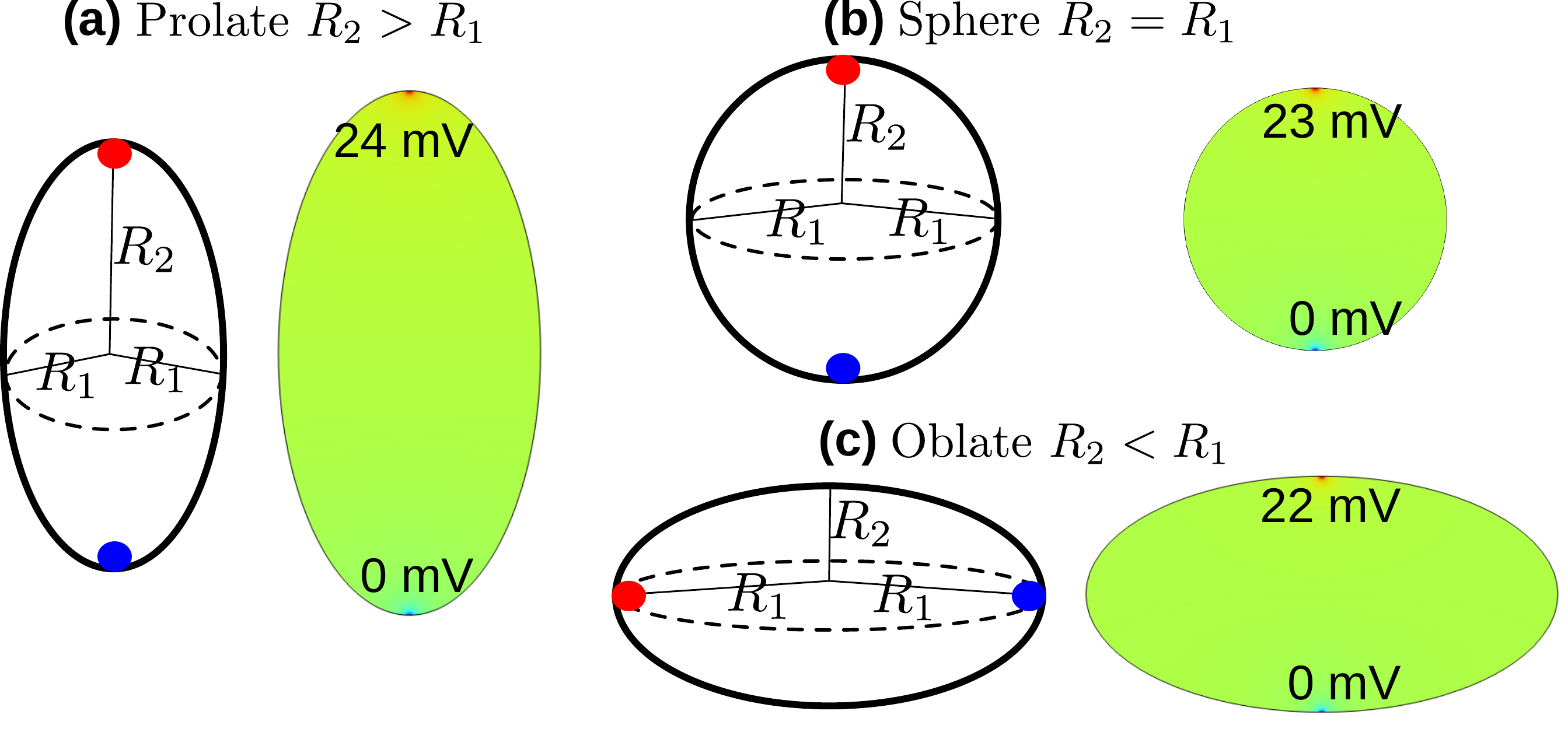}
\caption{\label{fig:fig1} \textbf{Spheroid geometries.} \textbf{(a)} Prolate spheroid with $R_1 = 0.5\,\mu{\rm m}$, $R_2 = 1\,\mu{\rm m}$ and influx/efflux at the North/South Poles. \textbf{(c)} Perfect sphere with $R_1 = R_2 = 500$ nm. \textbf{(c)} Oblate spheroid with $R_1 = 1\,\mu$m and $R_2 = 0.5\,\mu$m. 2-D voltage maps obtained with COMSOL \cite{comsol} for a two-charge model with $z_1 = +1$, $z_2 = -1$, $I = 100$ pA, $C_1 = C_2 = 100$ mM, $A_1 = A_2 = 10$ nm and other electro-diffusion parameters taken from Table \ref{table:param}.}
\end{figure}

%%%%%%%%%%%%%%%%%%%%%%%%%%%%%%%%%%%%%%%%%%%%%%%%%%%%%%%%%%%%%%%%%%%
% Make analogy between spheroid domains and synaptic current in head compartment of dendritic spines.
In this section we perform full numerical simulations of the PNP electro-diffusion model defined in Eq.~\eqref{eq:NP}-\eqref{eq:Poiss}, focusing on the $n=2,\,3$ ionic species case with results presented in terms of original dimensional variables. We also consider a class of spheroid domains $\Omega$ (Fig.~\ref{fig:fig1}) consisting of ellipsoid domains with azymuthal symmetry and axis described in the $x$, $y$ and $z$ directions by the set $(R_1,R_1,R_2)$. The case $R_1 < R_2$ corresponds to a prolate spheroid (Fig.~\ref{fig:fig1}\textbf{(a)}), while $R_1 > R_2$ yields an oblate spheroid (Fig.~\ref{fig:fig1}\textbf{(c)}) and with $R_1 = R_2$ corresponding to a perfect sphere (Fig.~\ref{fig:fig1}\textbf{(b)}). We use the azymuth $\phi$ and the colatitude $\theta$ to parameterize the domain boundary as
\beq
\p\Omega =\left\{ \left( \left. R_1\cos(\phi)\sin(\theta), R_1\sin(\phi)\sin(\theta), R_2\cos(\theta) \right) \right| 0 \leq \phi < 2\pi,\, 0 \leq \theta \leq \pi \right\}\,,
\eeq
but due to azymuthal invariance the mean curvature only depends on the colatitude $\theta$,
\beq
H(\theta) = \frac{R_2(2R_1^2 + (R_2^2 - R_1^2)\sin^2(\theta))}{2R_1(R_1^2 + (R_2^2 - R_1^2)\sin^2(\theta))^{3/2}}\,,
\eeq
which reduces to $H = 1/R_1$ when $R_1 = R_2$. The characteristic length-scale of the dimensional domain $\Omega$ is further defined as $R = \max\left\{R_1,\,R_2\right\}$. \\
Here the simulations shown in Fig.~\ref{fig:fig1} apply to voltage propagation within head compartments of dendritic spines following stimulation by a synaptic current \cite{yuste2010dendritic}, with these results confirming that these compartments are mostly equipotential \cite{cartailler2018neuron,paquin2024}.
%%%%%%%%%%%%%%%%%%%%%%%%%%%%%%%%%%%%%%%%%%%%%%%%%%%%%%%%%%%%%%%%%%%
%%%%%%%%%%%%%%%%%%%%%%%%%%%%%%%%%%%%%%%%%%%%%%%%%%%%%%%%%%%%%%%%%%%
%%%%%%%%%%%%%%%%%%%%%%%%%%%%%%%%%%%%%%%%%%%%%%%%%%%%%%%%%%%%%%%%%%%
\subsection{Comparing the two different parameter regimes}
\begin{figure}[!ht]
\centering
\includegraphics[width=0.66\linewidth]{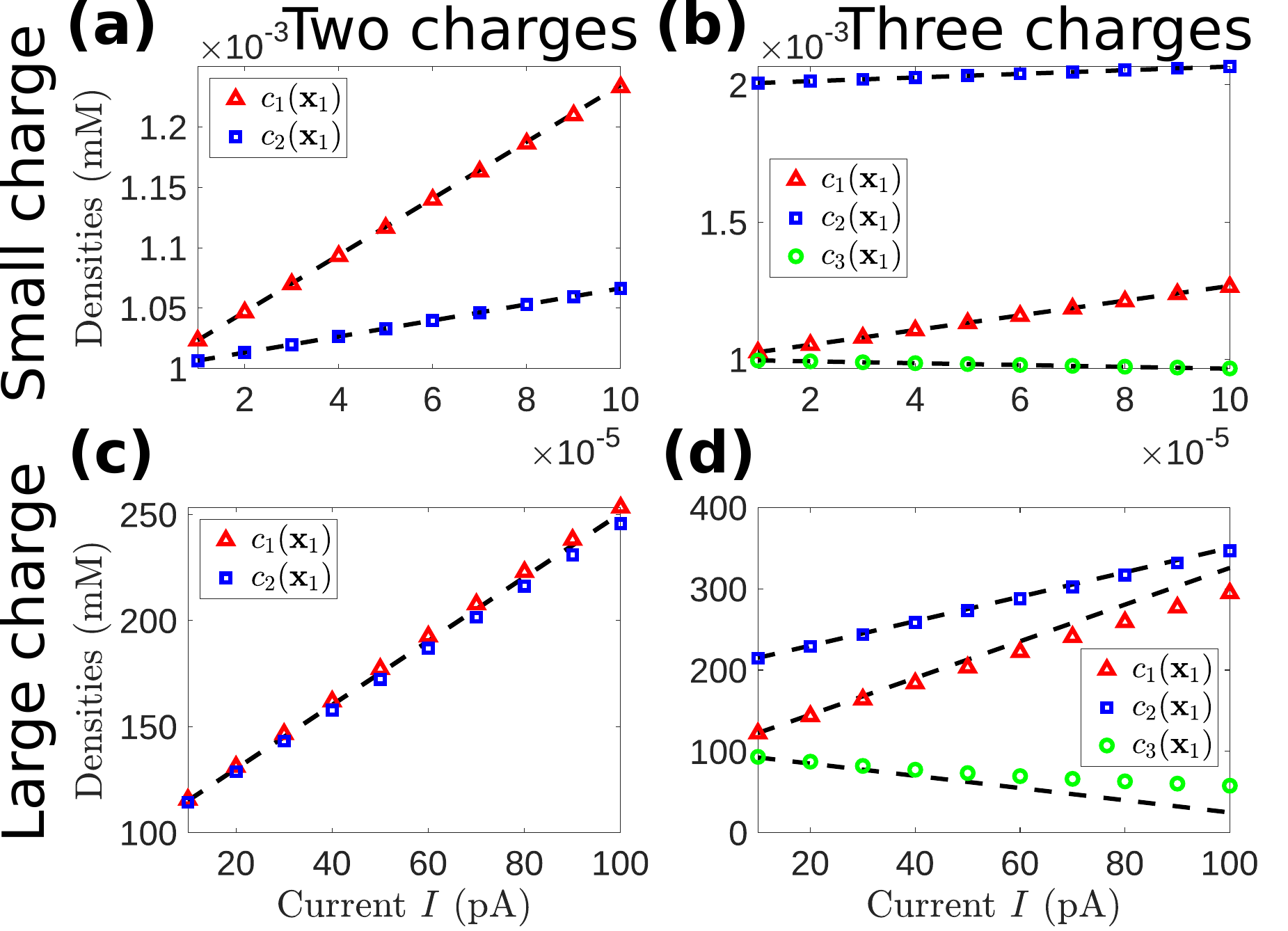}
\caption{\label{fig:ionic} \textbf{Current-concentrations relations.} Ionic densities versus influx current for $n=2$ \textbf{(a)} and $n=3$ \textbf{(b)} charges when the Debye length is of the order of the domain length-scale. Same plots in \textbf{(c)} and \textbf{(d)} but with the Debye length that is much smaller than the domain length-scale. The black dashed curves indicate ionic densities asymptotic solutions Eq.~\eqref{eq:alpha_small_c1}-\eqref{eq:alpha_small_ci} ($\alpha \sim O(1)$) and Eq.~\eqref{eq:alpha_large_c1}-\eqref{eq:alpha_large_ci} ($\alpha \gg 1$), while the symbols (triangles, squares and circles) show COMSOL \cite{comsol} simulation results. Ionic valences and background densities are as described in table \ref{table:regimes_z1_1}, with $\Omega$ a ball of radius $R=500$ nm. Parameters are $A_1 = A_2 = 10$ nm, others from Table \ref{table:param}.}
\end{figure}

\begin{figure}[!ht]
\centering
\includegraphics[width=0.66\linewidth]{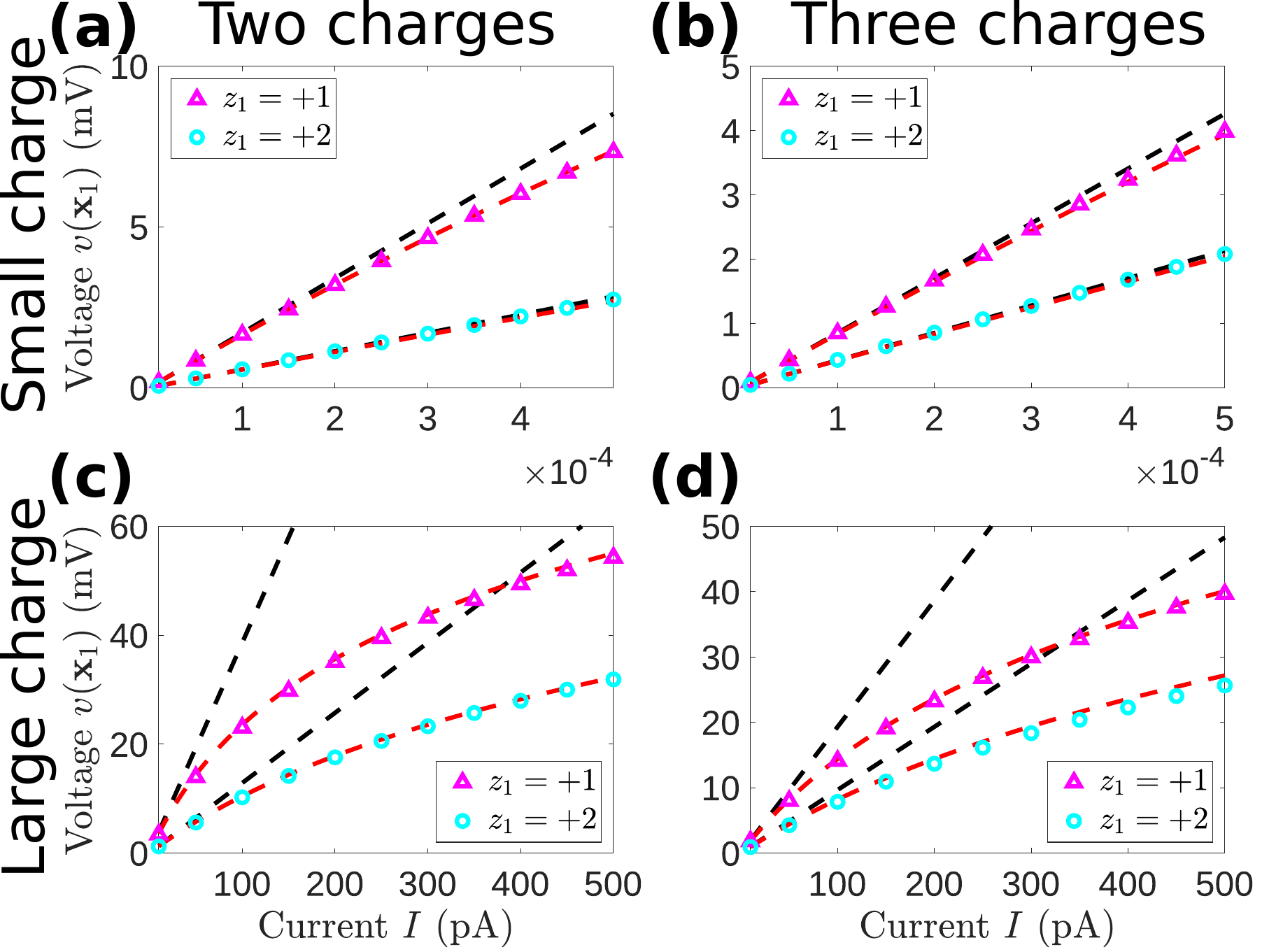}
\caption{\label{fig:voltage} \textbf{Current-voltage relations.} Current-voltage relations for $n=2$ \textbf{(a)} and $n=3$ \textbf{(b)} charges when the Debye length is of the order of the domain length-scale. Same plots in \textbf{(c)} and \textbf{(d)} but with the Debye length that is much smaller than the domain length-scale. Here we compare valences from Tables \ref{table:regimes_z1_1} ($z_1 = +1$) and \ref{table:regimes_z1_2} ($z_1 = +2$), with $\Omega$ a ball of radius $R=500$ nm. Parameters are $A_1 = A_2 = 10$ nm, others from Table \ref{table:param}.}
\end{figure}

%%%%%%%%%%%%%%%%%%%%%%%%%%%%%%%%%%%%%%%%%%%%%%%%%%%%%%%%%%%%%%%%%%%
We compare steady-state ionic and voltage dynamics for the regimes $\alpha \sim O(1)$ and $\alpha \gg 1$, i.e.~with the domain length-scale $R$ and the Debye length $\lambda_D$ satisfying either $\lambda_D \sim O(R)$ or $\lambda_D \ll R$. We consider two and three-charge electro-diffusion models with valences and background concentrations as described in Tables \ref{table:regimes_z1_1} and \ref{table:regimes_z1_2}. Ionic concentrations at the influx location are plotted against the current $I$ on a sphere of radius $R = 500$ nm (Fig.~\ref{fig:ionic}), with the regime $\alpha \sim O(1)$ leading to clear deviations from local electro-neutrality as expected. We observe a linear relation between the influx current and ionic densities at the influx location that is lost only at large current amplitude (Fig.~\ref{fig:ionic}\textbf{(d)}). Current-voltage I-V relations are plotted in Fig.~\ref{fig:voltage} where we find that the linear ohmic relations (black dashed curves) only agree when the current is small. In the highly nonlinear regime the voltage behavior follows a log formula (red dashed curves) given by
\beq\label{eq:I_V_log}
v(\x_1) = \frac{k_B\TT}{e}\log\left(1 + \frac{IF(A_1,A_2,R)}{\left(\sum_{k=1}^n z_k^2 C_k \right)\FF D_1 \pi A_2} \right)\,,
\eeq
with $F(A_1,A_2,R)$ defined as
\beq
F(A_1,A_2,R) =
\begin{cases}
\frac{\pi}{4} - \frac{H\left(\x_2\right)}{4}A_2\log\left(\frac{A_2}{R}\right) & \lambda_D \sim O(R) \\
\frac{A_2}{A_1} + \frac{\pi}{4} - \frac{A_2}{4} \left(H\left(\x_1\right)\log\left(\frac{A_1}{R}\right) + H\left(\x_2\right)\log\left(\frac{A_2}{R}\right)\right) & \lambda_D \ll R
\end{cases}\,.
\eeq
Our simulations also confirm that increasing the valence or adding more ions (in fact, any increase of the electrolyte conductivity $\sum_{k=1}^n z_k^2 C_k$) yield smaller voltage drop.
%%%%%%%%%%%%%%%%%%%%%%%%%%%%%%%%%%%%%%%%%%%%%%%%%%%%%%%%%%%%%%%%%%%
\begin{table}[!ht]
\centering
\caption{\label{table:regimes_z1_1} \textbf{Two and three-charge electro-diffusion models with $z_1 = +1$}}
\begin{tabular}{|l|l|l|l|}
\hline
\textbf{Charges} & \textbf{Electronic valence} & \textbf{Densities} (mM), $\lambda_D \sim O(R)$ & \textbf{Densities} (mM), $\lambda_D \ll R$ \\
\hline
$n=2$ & $z_1 = +1$, $z_2 = -1$ & $C_1 = 0.001$, $C_2 = 0.001$ & $C_1 = 100$, $C_2 = 100$ \\
\hline
$n=3$ & $z_1 = +1$, $z_2 = -1$, $z_3 = +1$ & $C_1 = C_3 = 0.001$, $C_2 = 0.002$ & $C_1 = C_3 = 100$, $C_2 = 200$ \\
\hline
\end{tabular}
\end{table}
%%%%%%%%%%%%%%%%%%%%%%%%%%%%%%%%%%%%%%%%%%%%%%%%%%%%%%%%%%%%%%%%%%%
%%%%%%%%%%%%%%%%%%%%%%%%%%%%%%%%%%%%%%%%%%%%%%%%%%%%%%%%%%%%%%%%%%%
\begin{table}[!ht]
\centering
\caption{\label{table:regimes_z1_2} \textbf{Two and three-charge electro-diffusion models with $z_1 = +2$}}
\begin{tabular}{|l|l|l|l|}
\hline
\textbf{Charges} & \textbf{Electronic valence} & \textbf{Densities} (mM), $\lambda_D \sim O(R)$ & \textbf{Densities} (mM), $\lambda_D \ll R$ \\
\hline
$n=2$ & $z_1 = +2$, $z_2 = -1$ & $C_1 = 0.001$, $C_2 = 0.002$ & $C_1 = 100$, $C_2 = 200$ \\
\hline
$n=3$ & $z_1 = +2$, $z_2 = -1$, $z_3 = +1$ & $C_1 = C_3 = 0.001$, $C_2 = 0.003$ & $C_1 = C_3 = 100$, $C_2 = 300$ \\
\hline
\end{tabular}
\end{table}
%%%%%%%%%%%%%%%%%%%%%%%%%%%%%%%%%%%%%%%%%%%%%%%%%%%%%%%%%%%%%%%%%%%
%%%%%%%%%%%%%%%%%%%%%%%%%%%%%%%%%%%%%%%%%%%%%%%%%%%%%%%%%%%%%%%%%%%
\begin{table}[!ht]
\centering
\caption{\label{table:param} \textbf{Electro-diffusion parameters.}}
\begin{tabular}{|l|c|l|}
\hline
\textbf{Parameter} & \textbf{Symbol} & \textbf{Value} \\
\hline
Diffusion coefficient of influx cation & $D_1$ & $200\,\mu$m$^2$s$^{-1}$ \\
Electron charge & $e$ & $1.60 \times 10^{-19}$ C \\
Avogadro constant & $N_A$ & $6.02 \times 10^{23}$ mol$^{-1}$ \\
Boltzmann constant & $k_B$ & $1.38 \times 10^{-23}$ J K$^{-1}$ \\
Temperature & $\TT$ & $298$ K \\
Relative permittivity & $\eps$ & $78.4$ \\
Vacuum permittivity & $\eps_0$ & $8.85\times 10^{-12}$ C V$^{-1}$m$^{-1}$ \\
Faraday constant & $\FF = eN_A$ & $96\,485.33$ C mol$^{-1}$ \\
Thermal voltage & $k_B\TT/e$ & $25.68$ mV \\
\hline
\end{tabular}
\end{table}
%%%%%%%%%%%%%%%%%%%%%%%%%%%%%%%%%%%%%%%%%%%%%%%%%%%%%%%%%%%%%%%%%%%
%%%%%%%%%%%%%%%%%%%%%%%%%%%%%%%%%%%%%%%%%%%%%%%%%%%%%%%%%%%%%%%%%%%
%%%%%%%%%%%%%%%%%%%%%%%%%%%%%%%%%%%%%%%%%%%%%%%%%%%%%%%%%%%%%%%%%%%
%%%%%%%%%%%%%%%%%%%%%%%%%%%%%%%%%%%%%%%%%%%%%%%%%%%%%%%%%%%%%%%%%%%
%%%%%%%%%%%%%%%%%%%%%%%%%%%%%%%%%%%%%%%%%%%%%%%%%%%%%%%%%%%%%%%%%%%
%%%%%%%%%%%%%%%%%%%%%%%%%%%%%%%%%%%%%%%%%%%%%%%%%%%%%%%%%%%%%%%%%%%
\subsection{Domain geometry and membrane curvature}
\begin{figure}[!ht]
\centering
\includegraphics[width=0.66\linewidth]{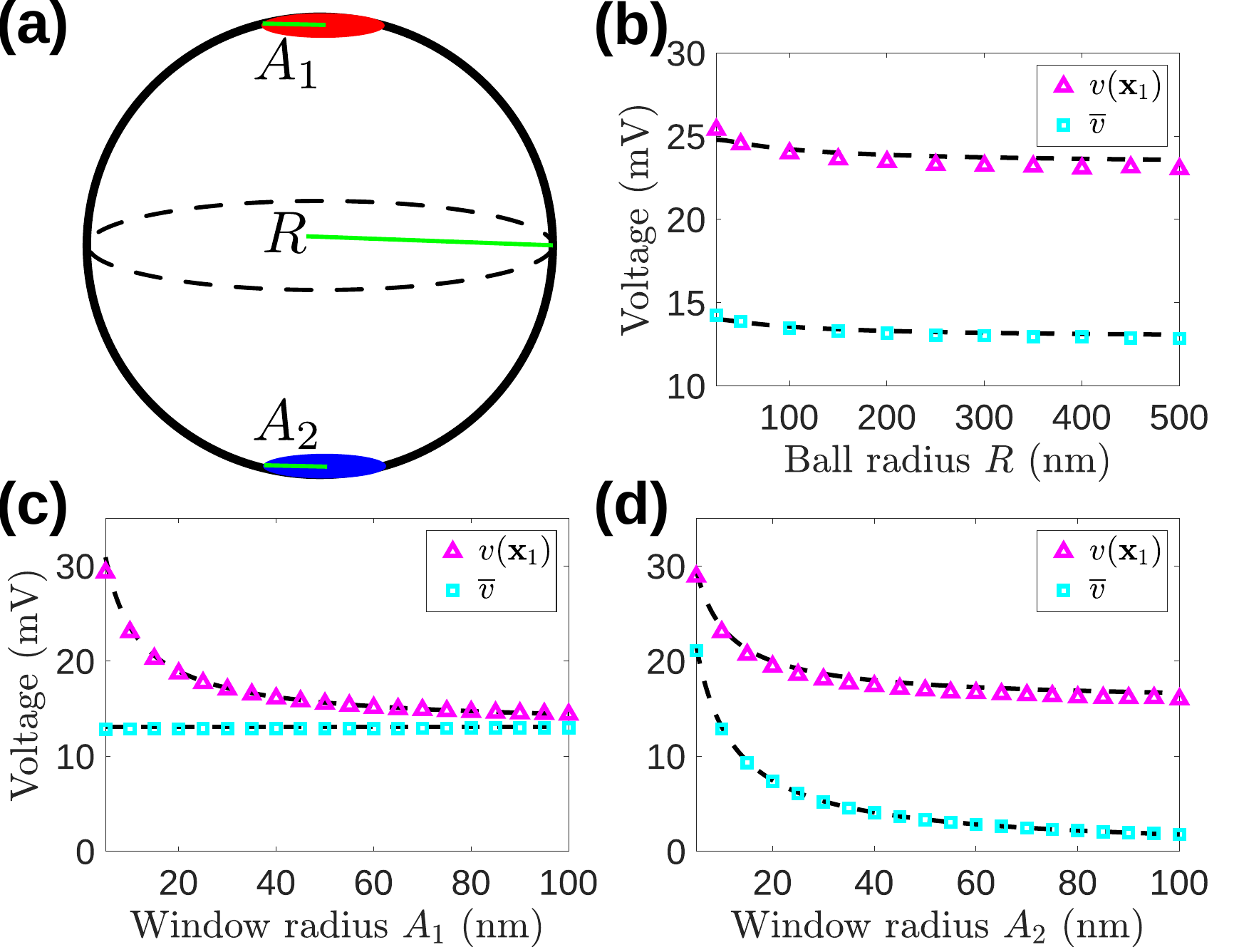}
\caption{\label{fig:win_size} \textbf{Window channels size.} Influx location voltage $v(\x_1)$ and voltage domain average $\overline{v}$ as a function of the ball radius $R$ with $A_1 = A_2 = 10$ nm \textbf{(a)}, the influx window radius $A_1$ with $R=500$ nm and $A_2 = 10$ nm \textbf{(b)}, and the exit window radius $A_2$ with $R=500$ nm and $A_1 = 10$ nm \textbf{(c)}. Black dashed curves correspond to asymptotic solutions Eq.~\eqref{eq:I_V_log}-\eqref{eq:I_V_barlog} and symbols (triangles and squares) are COMSOL \cite{comsol} simulation results. Parameters are $z_1 = +1$, $z_2 = -1$, $C_1 = C_2 = 100$ mM, $I = 100$ pA and others taken from Table \ref{table:param}.}
\end{figure}

\begin{figure}[!ht]
\centering
\includegraphics[width=\linewidth]{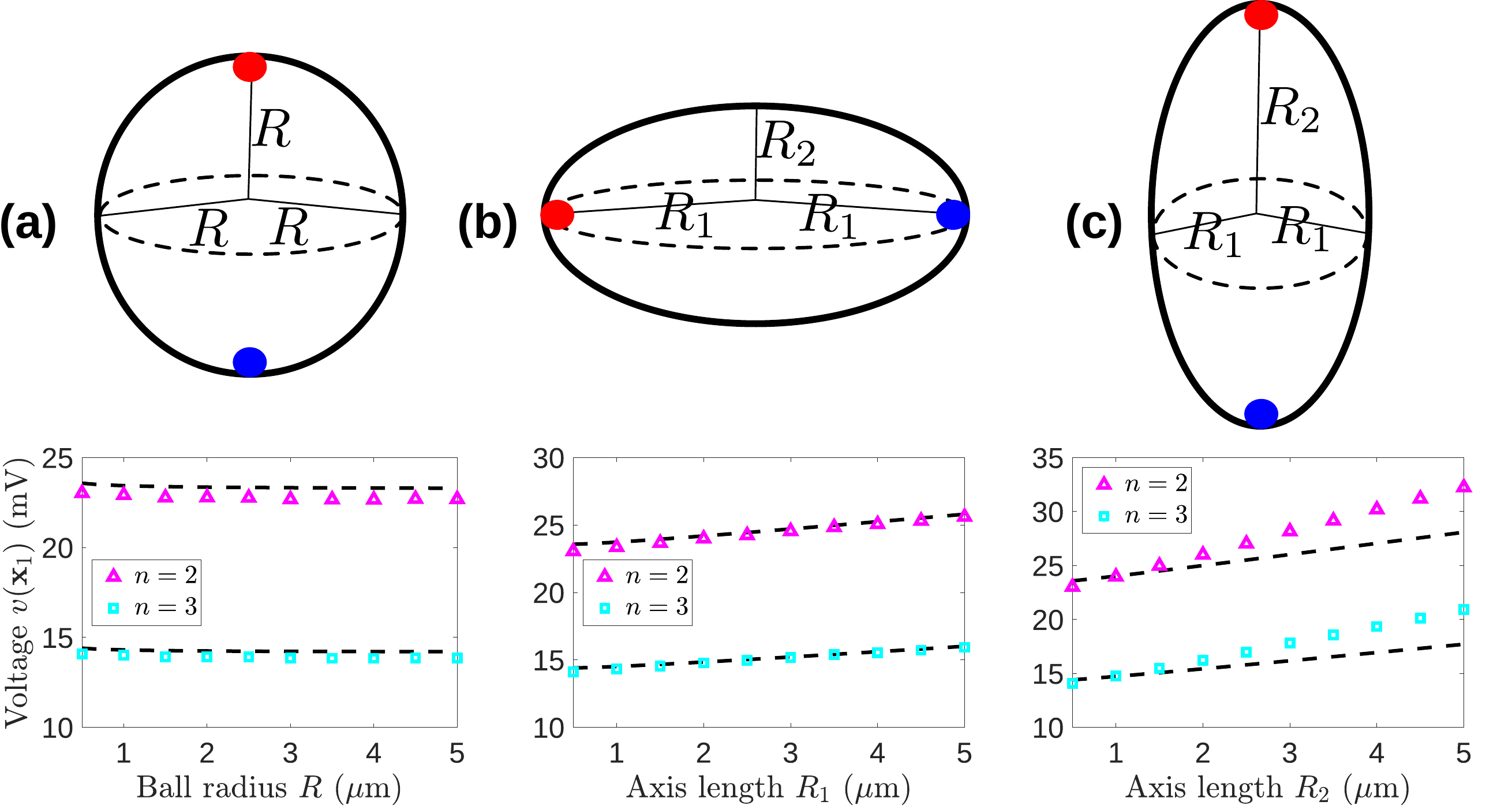}
\caption{\label{fig:mem_curv} \textbf{Voltage dynamics modulated by mean membrane curvature.} \textbf{(a) Sphere:} Voltage vs radius $R$ with influx/efflux at the North/South Poles. \textbf{(b) Oblate:} Voltage vs spheroid axis $R_1$ with $R_2 = 500$ nm and influx/efflux on the Equator (azymuth $\phi_1 = 0,\,\phi_2 = \pi$). \textbf{(c) Prolate:} Voltage vs spheroid axis $R_2$ with $R_1 = 500$ nm and influx/efflux at the North/South Poles. Black dashed curves are numerically evaluated using Eq.~\eqref{eq:I_V_log} and symbols (triangles and squares) are COMSOL \cite{comsol} simulation results for two and three-charge models with $z_1 = +1$ (cf.~Table \ref{table:regimes_z1_1}, large charge regime). Parameters are $I=100$ pA, $A_1 = A_2 = 10$ nm with others taken from Table \ref{table:param}.}
\end{figure}

%%%%%%%%%%%%%%%%%%%%%%%%%%%%%%%%%%%%%%%%%%%%%%%%%%%%%%%%%%%%%%%%%%%
We proceed by investigating how the domain, channels size and mean membrane curvature can affect voltage dynamics (Fig.~\ref{fig:win_size}-\ref{fig:mem_curv}) in the regime $\alpha \gg 1$. This is first shown with the simple case of two monovalent ionic species on a ball of radius $R$, where we vary the ball radius and the narrow radii $A_1$ and $A_2$ (Fig.~\ref{fig:win_size}). In addition to the voltage at the influx location, as given by relation Eq.~\eqref{eq:I_V_log}, we compare the average domain voltage with the log relation below
\beq\label{eq:I_V_barlog}
\overline{v} = \frac{k_B\TT}{e}\log\left(1 + \frac{I}{\left(\sum_{k=1}^n z_k^2 C_k \right)\FF D_1 \pi A_2} \left( \frac{\pi}{4} - \frac{H\left(\x_2\right)}{4}A_2\log\left(\frac{A_2}{R}\right) \right) \right) \,,
\eeq
with mean membrane curvature satisfying $H(\x) = 1/R$ since $\Omega$ is a ball. These asymptotic relations suggest that when $A_1,\,A_2 \ll R$, any further increase of the domain radius has little effect on the voltage. Interestingly the size of the influx window has no effect on the average voltage $\overline{v}$. As in \cite{paquin2024} both numerical and asymptotic solutions suggest that the mean membrane curvature affects voltage dynamics much more strongly than the Euclidean distance between channels (Fig.~\ref{fig:mem_curv}). This is seen by comparing voltage drop on a ball (Fig.~\ref{fig:mem_curv}\textbf{(a)}) versus oblate (Fig.~\ref{fig:mem_curv}\textbf{(b)}) and prolate (Fig.~\ref{fig:mem_curv}\textbf{(c)}) spheroid domains. In all cases the Euclidean distance between the influx and the efflux is the same, what changes is the local mean curvature at the influx/efflux locations. For the sphere with $R_1 = R_2$ and $H(\x) = \frac{1}{R_1}$ the curvature behaves reciprocally with the radius (Fig.~\ref{fig:mem_curv}\textbf{(a)}), while on a prolate spheroid with $\theta_1 = 0,\,\theta_2 = \pi$ the mean curvature satisfies $H(\x_1) = H(\x_2) = R_2/R_1^2$ and increases linearly with the axis $R_2$ (Fig.~\ref{fig:mem_curv}\textbf{(c)}). On the oblate spheroid with influx/efflux along the Equator (Fig.~\ref{fig:mem_curv}\textbf{(b)}), the mean curvature function evaluates as
\beq
H(\x_1) = H(\x_2) = \frac{R_1}{2}\left( \frac{1}{R_1^2} + \frac{1}{R_2^2} \right)\,,
\eeq
and we also find that stretching $R_1$ yields a larger voltage drop.
%%%%%%%%%%%%%%%%%%%%%%%%%%%%%%%%%%%%%%%%%%%%%%%%%%%%%%%%%%%%%%%%%%%%%%%%%%%%%%%%%
%%%%%%%%%%%%%%%%%%%%%%%%%%%%%%%%%%%%%%%%%%%%%%%%%%%%%%%%%%%%%%%%%%%%%%%%%%%%%%%%%
%%%%%%%%%%%%%%%%%%%%%%%%%%%%%%%%%%%%%%%%%%%%%%%%%%%%%%%%%%%%%%%%%%%%%%%%%%%%%%%%%
%%%%%%%%%%%%%%%%%%%%%%%%%%%%%%%%%%%%%%%%%%%%%%%%%%%%%%%%%%%%%%%%%%%%%%%%%%%%%%%%%
%%%%%%%%%%%%%%%%%%%%%%%%%%%%%%%%%%%%%%%%%%%%%%%%%%%%%%%%%%%%%%%%%%%%%%%%%%%%%%%%%
\section{Discussion}\label{sec:perspec}
%%%%%%%%%%%%%%%%%%%%%%%%%%%%%%%%%%%%%%%%%%%%%%%%%%%%%%%%%%%%%%%%%%%%%%%%%%%%%%%%%
In this article we found I-V relations inside nanocellular electrolytes containing multiple ionic species. The voltage is modeled using electro-diffusion theory formulated with the PNP equations. After normalizing the main variables (dimensional scaling) in Section \S \ref{sec:model}, we find two key parameters controlling how voltage spreads and drops within the nano-domain: the influx current $J$ and the ratio $\alpha = R/\lambda_D$ between the domain length-scale and the Debye length. A spatial rescaling with the Debye length then allows us to perform the computation based on a regular asymptotic expansion in terms of the small ratio $J/\alpha$ and solve for linearized ionic and voltage solutions at $O(J/\alpha)$ using Green's function methods in \S \ref{sec:asym}. \\
The main advantage of this expansion is to allow $\alpha$ to be of order one (with $R \sim \lambda_D$) or to be large with $\alpha \gg 1$ ($R < \lambda_D$). While the latter case leads to no significant deviation from electro-neutrality, we find that when $\alpha \sim O(1)$ ionic concentrations at the influx location deviate from local electro-neutrality with the maximal voltage drop well-approximated by the voltage domain average. The I-V formula derived here further reveals the role of the local mean membrane curvature in modulating ionic and voltage dynamics. Finally, numerical simulations illustrate how the curvature of three-dimensional prolate and oblate spheroid domains affect voltage and concentration distributions (Section \S \ref{sec:numerics}). \\
It would be interesting to extend the current asymptotic theory to transient dynamical behaviors following a time-dependent influx. This would allow us to determine how recovery response time-scales are affected by membrane curvature and by the geometry of subcellular nanodomains.
%%%%%%%%%%%%%%%%%%%%%%%%%%%%%%%%%%%%%%%%%%%%%%%%%%%%%%%%%%%%%%%%%%%%%%%%%%%%%%%%%
\section*{Acknowledgements}
%%%%%%%%%%%%%%%%%%%%%%%%%%%%%%%%%%%%%%%%%%%%%%%%%%%%%%%%%%%%%%%%%%%%%%%%%%%%%%%%%
\noindent F.~P.-L.~gratefully acknowledges the support from the Natural Sciences and Engineering Research Council of Canada (NSERC) (Award No 578183-2023) and from the Fondation ARC (Award No ARCPDF12020020001505) through postdoctoral fellowships. D.~H.~was supported by the European Research Council (ERC) under the European Union’s Horizon 2020 Research and Innovation Program (grant agreement No 882673), and ANR AstroXcite.
%%%%%%%%%%%%%%%%%%%%%%%%%%%%%%%%%%%%%%%%%%%%%%%%%%%%%%%%%%%%%%%%%%%%%%%%%%%%%%%%%
%%%%%%%%%%%%%%%%%%%%%%%%%%%%%%%%%%%%%%%%%%%%%%%%%%%%%%%%%%%%%%%%%%%%%%%%%%%%%%%%%
\begin{appendix}
%%%%%%%%%%%%%%%%%%%%%%%%%%%%%%%%%%%%%%%%%%%%%%%%%%%%%%%%%%%%%%%%%%%%%%%%%%%%%%%%%
%%%%%%%%%%%%%%%%%%%%%%%%%%%%%%%%%%%%%%%%%%%%%%%%%%%%%%%%%%%%%%%%%%%%%%%%%%%%%%%%%
\section{Computing boundary integrals with the Neumann Green's function}\label{sec:app}
%%%%%%%%%%%%%%%%%%%%%%%%%%%%%%%%%%%%%%%%%%%%%%%%%%%%%%%%%%%%%%%%%%%%%%%%%%%%%%%%%
%%%%%%%%%%%%%%%%%%%%%%%%%%%%%%%%%%%%%%%%%%%%%%%%%%%%%%%%%%%%%%%%%%%%%%%%%%%%%%%%%
\subsection{The Neumann Green's function}\label{sec:green}
%%%%%%%%%%%%%%%%%%%%%%%%%%%%%%%%%%%%%%%%%%%%%%%%%%%%%%%%%%%%%%%%%%%%%%%%%%%%%%%%%
%%%%%%%%%%%%%%%%%%%%%%%%%%%%%%%%%%%%%%%%%%%%%%%%%%%%%%%%%%%%%%%%%%%%%%%%%%%%%%%%%
We start with the domain $\tilde{\Omega}$, of $O(1)$ characteristic length-scale, on which we can introduce the Neumann Green's function $\ggreen$ solution of
\begin{equation}
\Delta \ggreen = \frac{1}{\abs{\tilde{\Omega}}}, \quad \tx \in \tilde{\Omega}, \quad \frac{\p \ggreen}{\partial \n} = \delta(\tx - \ty), \quad \tx \in \p \tilde{\Omega}, \quad \int_{\tilde{\Omega}} \ggreen d\tx = 0, \quad \ty \in \p \tilde{\Omega}\,,
\end{equation}
that behaves near the singular diagonal $\tx = \ty$ as
\begin{equation}
\ggreen \approx \frac{1}{2\pi \norm{\tx - \ty}} - \frac{\tilde{H}(\ty)}{4 \pi} \log(\norm{\tx - \ty}) + \tilde{R}_s(\tx;\ty)\,, \quad \text{with} \quad 0 < \norm{\tx-\ty} \ll 1\,,
\end{equation}
with mean boundary curvature function $\tilde{H}(\ty)$ and where $\tilde{R}_s(\tx;\ty)$ is the regular part, which can be further approximated as
\begin{equation}
\tilde{R}_s(\tx;\ty) \approx \tilde{R}_s(\ty;\ty) + O(\norm{\tx - \ty})\,.
\end{equation}
However on the stretched domain $\Omega_{\bxi}$ with $\bxi = \alpha\tx$ and $\bta = \alpha\ty$, we can work with
\beq
\GGREEN = \frac{1}{\alpha}\tilde{G}_s\left(\frac{\bxi}{\alpha};\frac{\bta}{\alpha}\right)\,,
\eeq
which satisfies
\begin{equation}\label{eq:green_bxi}
\Delta \GGREEN = \frac{1}{\abs{\Omega_{\bxi}}}, \quad \bxi \in \Omega_{\bxi}, \quad \frac{\p \GGREEN}{\p \n} = \delta(\bxi - \bta), \quad \bxi \in \p \Omega_{\bxi}, \quad \int_{\Omega_{\bxi}}\GGREEN d\bxi = 0, \quad \bta \in \p \Omega_{\bxi}\,,
\end{equation}
and behaves near the singular diagonal $\bxi = \bta$ as
\begin{equation}
\GGREEN \approx \frac{1}{2 \pi \norm{\bxi - \bta}} - \frac{\HH(\bta)}{4 \pi} \log\left(\norm{\bxi - \bta}\right) + \RR_s(\bxi;\bta), \quad 0 < \norm{\bxi - \bta} \ll \alpha\,,
\end{equation}
where $\HH(\bta)$ and $\RR_s(\bxi;\bta)$ are defined as
\beq
\HH(\bta) = \frac{1}{\alpha}\tilde{H}\left(\frac{\bta}{\alpha}\right), \quad \RR_s(\bxi;\bta) = \frac{1}{\alpha}\tilde{R}_s\left(\frac{\bxi}{\alpha};\frac{\bta}{\alpha}\right) + \frac{1}{4\pi\alpha}\tilde{H}\left(\frac{\bta}{\alpha}\right)\log(\alpha)\,,
\eeq
with $\RR_s(\bxi;\bta)$ that can be approximated as
\beq
\RR_s(\bxi;\bta) \approx \RR_s(\bta;\bta) + O\left(\frac{\norm{\bxi - \bta}}{\alpha}\right)\,.
\eeq
%%%%%%%%%%%%%%%%%%%%%%%%%%%%%%%%%%%%%%%%%%%%%%%%%%%%%%%%%%%%%%%%%%%%%%%%%%%%%%%%%
\subsection{Weber solution from electrostatics}\label{sec:weber}
%%%%%%%%%%%%%%%%%%%%%%%%%%%%%%%%%%%%%%%%%%%%%%%%%%%%%%%%%%%%%%%%%%%%%%%%%%%%%%%%%
Here we provide a justification for why the exit flux can be approximated by the classical Weber solution \cite{crank1975}. We consider
\beq\label{eq:outer}
\Delta \PP(\bxi) - \PP(\bxi) = 0, \quad \bxi \in \Omxi \,,
\eeq
along with absorbing boundary $\PP(\bxi) = 0$ for $\bxi\in\p\oxio$. Following \cite{gomez2015} we can use the small radius $a_2\alpha$ to introduce a set of local cartesian coordinates $(\eta,s_1,s_2)$ with which we can map the neighborhood near $\p\oxio$ to the infinite half-space in 3-D. This mapping relies on the expansion of the Laplacian as
\begin{equation}
\Delta_{\bxi} = \frac{1}{\left(\alpha a_2\right)^2} \Delta_{(\eta,s_1,s_2)} + O\left(\frac{1}{\alpha a_2}\right) = \frac{1}{\left(\alpha a_2\right)^2}\left(\frac{\p^2}{\p\eta^2} + \frac{\p^2}{\p s_1^2} + \frac{\p^2}{\p s_2^2}\right) + O\left(\frac{1}{\alpha a_2}\right) \,,
\end{equation}
where $\Delta_{(\eta,s_1,s_2)}$ is the Laplacian expressed in Cartesian coordinates. By redefining $\PP(\bxi)$ as $P(\eta,s_1,s_2)$ Eq.~\eqref{eq:outer} gets transformed as
\beq
\left(\frac{1}{\left(\alpha a_2\right)^2} \Delta_{(\eta,s_1,s_2)} + O\left(\frac{1}{\alpha a_2}\right)\right) P(\eta,s_1,s_2) - P(\eta,s_1,s_2) = 0,
\eeq
and thus at leading-order we obtain
\beq\label{eq:inner}
\Delta_{(\eta,s_1,s_2)} P(\eta,s_1,s_2) = 0, \quad \left. P\right|_{\eta = 0} = 0, \quad \sqrt{s_1^2 + s_2^2} < 1, \quad \left. \frac{\p P}{\p \eta}\right|_{\eta = 0} = 0, \quad \sqrt{s_1^2 + s_2^2} > 1\,,
\eeq
$P(\eta,s_1,s_2) \to P_0$ as far-field condition $\sqrt{\eta^2 + s_1^2 + s_2^2} \to \infty$. The solution to Eq.~\eqref{eq:inner} is given by
\beq
P(\eta,s_1,s_2) = P_0\left(1 - \frac{2}{\pi}\int_0^\infty e^{-\eta m}J_0\left(m\sqrt{s_1^2 + s_2^2}\right)\sin(m)\frac{dm}{m}\right)\,,
\eeq
where $J_0$ is the usual regular Bessel function of order 0. By then computing the flux we get
\beq
\left.\frac{\p P}{\p\eta}\right|_{\eta = 0} = \frac{2P_0}{\pi} \frac{1}{\sqrt{1 - (s_1^2+s_2^2)}}, \quad \sqrt{s_1^2 + s_2^2} < 1\,,
\eeq
and then when transformed back to the coordinates $\bxi$ this becomes
\beq
\frac{\p \PP(\bxi)}{\p \bxi} \sim \frac{1}{\sqrt{\left(\alpha a_2\right)^2 - \|\bxi-\bxi_2\|^2}}, \quad \bxi \in \p\oxio\,.
\eeq
We can treat similarly the equations for $\SS_{ij}$ and $\VV$.
%%%%%%%%%%%%%%%%%%%%%%%%%%%%%%%%%%%%%%%%%%%%%%%%%%%%%%%%%%%%%%%%%%%%%%%%%%%%%%%%%
\subsection{Computing boundary flux integrals}\label{sec:integrals}
%%%%%%%%%%%%%%%%%%%%%%%%%%%%%%%%%%%%%%%%%%%%%%%%%%%%%%%%%%%%%%%%%%%%%%%%%%%%%%%%%
We now derive approximation for the singular integrals from Section \S \ref{sec:asym} which involve the Neumann Green's function. We first calculate the boundary integrals
\beq
\int_{\p\Omega_{\bxi_1}} \GG_s(\bxi;\bxi_1) d\bxi \quad \text{and} \quad \int_{\p\Omega_{\bxi_2}} \frac{\GG_s(\bxi;\bxi_2)}{\sqrt{\alpha a_2^2 - \|\bxi-\bxi_2\|^2}} d\bxi\,,
\eeq
by defining $\rho = \|\bxi-\bxi_i\|$ and using polar coordinates. We obtain that
\begin{align}
\int_{\p\Omega_{\bxi_1}} \GG_s(\bxi;\bxi_1) d\bxi &= 2\pi\int_0^{a_1\alpha} \left( \frac{1}{2\pi \rho} - \frac{\HH(\bxi_1)}{4\pi}\log\left(\rho\right) + \RR_s(\bxi_1;\bxi_1) + O\left(\frac{\rho}{\alpha}\right) \right) \rho d\rho\,, \nonumber \\
&= \alpha a_1 - \frac{\HH(\bxi_1)}{2}\alpha^2 a_1^2\left(\frac{1}{2}\log(\alpha a_1) - \frac{1}{4}\right) + \pi \alpha^2 a_1^2\RR_s(\bxi_1;\bxi_1) + O\left(\alpha a_1^3\right)\,, \nonumber \\
&= \alpha a_1 f(a_1)\,, \nonumber
\end{align}
with $f(a_1)$ defined as
\begin{equation}
f(a_1) = 1 - \alpha\frac{\HH(\bxi_1)}{4}a_1\log(a_1) + \alpha a_1\left( \pi\RR_s(\bxi_1;\bxi_1) + \frac{\HH(\bxi_1)}{4}\left(\frac{1}{2} - \log(\alpha)\right)\right) + O(a_1^2)\,,
\end{equation}
but then on the domain $\tilde{\Omega}$ with $O(1)$ length-scale this yields,
\beq
f(a_1) = 1 - \frac{\tilde{H}(\tx_1)}{4}a_1\log(a_1) + a_1\left(\pi \tilde{R}_s(\tx_1;\tx_1) + \frac{\tilde{H}(\tx_1)}{8}\right) + O(a_1^2)\,.
\eeq
We then proceed similarly with the integral involving the classical Weber solution, and evaluate
\begin{align*}
\int_{\p\Omega_{\bxi_2}} \frac{\GG_s(\bxi;\bxi_2)}{\sqrt{\alpha a_2^2 - \|\bxi-\bxi_2\|^2}} d\bxi &= 2\pi \int_0^{\alpha a_2} \left(\frac{1}{2\pi \rho} - \frac{\HH(\bxi_2)}{4\pi}\log\left(\rho\right) + \RR_s(\bxi_2;\bxi_2) + O\left(\frac{\rho}{\alpha}\right) \right) \frac{\rho d\rho}{\sqrt{\left(\alpha a_2\right)^2 - \rho^2}}\,, \\
&= \frac{\pi}{2} - \alpha\frac{\HH(\bxi_2)}{2} a_2\left( \log(\alpha a_2) + \log(2) - 1 \right) + 2\pi a_2\alpha \RR_s(\bxi_2;\bxi_2) + O\left(a_2^2\right)\,, \\
&= \frac{\pi}{2}g(a_2)\,,
\end{align*}
where $g(a_2)$ is defined as
\begin{equation}
g(a_2) = 1 - \alpha \frac{\HH(\bxi_2)}{\pi}a_2\log(a_2) + \alpha a_2 \left(4\RR_s(\bxi_2;\bxi_2) + \frac{\HH(\bxi_2)}{\pi}\left(1-\log(2) - \log(\alpha)\right) \right) + O(a_2^2)\,,
\end{equation}
but then on the domain $\tilde{\Omega}$ with $O(1)$ length-scale this yields
\begin{equation}
g(a_2) = 1 - \frac{\tilde{H}(\tx_2)}{\pi}a_2\log(a_2) + a_2 \left(4\tilde{R}_s(\tx_2;\tx_2) +  \frac{\tilde{H}(\tx_2)}{\pi}(1-\log(2))\right) + O(a_2^2)\,.
\end{equation}
By then assuming that the narrow windows are well-spaced we get the approximation below for the mixed integral terms
\begin{align}
&\int_{\p\Omega_{\bxi_1}} \GG_s(\bxi;\bxi_2) d\bxi \approx \pi \left(a_1\alpha\right)^2 \GG_s(\bxi_1;\bxi_2), \\
&\int_{\p\Omega_{\bxi_2}} \frac{\GG_s(\bxi;\bxi_1)}{\sqrt{\alpha a_2^2 - \|\bxi-\bxi_2\|^2}} d\bxi \approx 2\pi a_2\alpha \GG_s(\bxi_2;\bxi_1)\,.
\end{align}
%%%%%%%%%%%%%%%%%%%%%%%%%%%%%%%%%%%%%%%%%%%%%%%%%%%%%%%%%%%%%%%%%%%%%%%%%%%%%%%%%
To calculate the last integral in Eq.~\eqref{eq:greenid_sub_pp_1}-\eqref{eq:greenid_sub_pp_2} we proceed by splitting the domain into three subregions,
\begin{equation*}\label{eq:int_3D}
\int_{\Omxi} \GG_s(\bxi;\bxi_j)\PP(\bxi) d\bxi = \left(\int_{B\left(\bxi_1,\alpha a_1\right)} + \int_{B\left(\bxi_2,\alpha a_2\right)} + \int_{\Omxi\backslash\left\{B\left(\bxi_1,\alpha a_1\right) \cup B\left(\bxi_2,\alpha a_2\right) \right\}} \right)\GG_s(\bxi;\bxi_j)\PP(\bxi)d\bxi, \quad j=1,2\,,
\end{equation*}
where $B\left(\bxi_j,\alpha a_j\right)$ for $j=1,2$ are half-spheres of radius $\alpha a_j$ centered in $\bxi_j$. Outside of these subregions the sum $\PP(\bxi)$ is constant, yielding
\begin{equation*}
\int_{\Omxi\backslash\left\{B\left(\bxi_1,\alpha a_1\right) \cup B\left(\bxi_2,\alpha a_2\right) \right\}} \GG_s(\bxi;\bxi_j)\PP(\bxi)d\bxi \approx \overline{\PP}\int_{\Omxi\backslash\left\{B\left(\bxi_1,\alpha a_1\right) \cup B\left(\bxi_2,\alpha a_2\right) \right\}} \GG_s(\bxi;\bxi_j) d\bxi \approx \PP \left[ O(1) \right]\,,
\end{equation*}
with its contribution that can be neglected, while for the two subregions we get
\begin{align*}
\int_{\Omega_{\bxi}} \GG_s(\bxi;\bxi_j)\PP(\bxi) d\bxi &\approx \left( \int_{B\left(\bxi_1,\alpha a_1\right)} + \int_{B\left(\bxi_2,\alpha a_2\right)} \right) \GG_s(\bxi;\bxi_j)\PP(\bxi) d\bxi\,, \\
&\approx \PP(\bxi_1)\int_{B\left(\bxi_1,\alpha a_1\right)} \GG_s(\bxi;\bxi_j) d\bxi + \PP(\bxi_2)\int_{B\left(\bxi_2,\alpha a_2\right)} \GG_s(\bxi;\bxi_j) d\bxi \\
&\approx \PP(\bxi_1)\int_{B\left(\bxi_1,\alpha a_1\right)} \GG_s(\bxi;\bxi_j) d\bxi
\end{align*}
due to the absorbing boundary condition $\PP(\bxi_2) = 0$. Here the windows are well-spaced, thus we calculate
\beq
\int_{B\left(\bxi_1,\alpha a_1\right)} \GG_s(\bxi;\bxi_2) d\bxi \approx \GG_s(\bxi_1;\bxi_2) \left| B\left(\bxi_1,\alpha a_1\right) \right| = \GG_s(\bxi_1;\bxi_2) \frac{2\pi}{3}\alpha^3 a_1^3\,,
\eeq
while around the singularity we get
\begin{align*}
\int_{B\left(\bxi_1,\alpha a_1\right)} \GG_s(\bxi;\bxi_1) d\bxi &= \int_{\frac{\pi}{2}}^\pi \int_0^{2\pi} \int_0^{\alpha a_1} \left( \frac{1}{2\pi\rho} - \frac{\HH(\bxi_1)}{4\pi}\log\left(\rho\right) + \RR_s(\bxi_1;\bxi_1) + O\left(\frac{\rho}{\alpha}\right) \right) \rho^2 \sin(\theta) d\rho d\phi d\theta\,, \\
&= \frac{\left(\alpha a_1\right)^2}{2} - \frac{\HH(\bxi_1)}{2}\alpha^3 a_1^3\left(\frac{1}{3}\log(\alpha a_1)  - \frac{1}{9}\right) + 2\pi\RR_s(\bxi_1;\bxi_1)\frac{\alpha^3 a_1^3}{3} + O\left(\alpha a_1^4\right)\,, \\
&= \frac{\left(\alpha a_1\right)^2}{2}m(a_1)\,,
\end{align*}
where $m(a_1)$ is an expansion given by
\begin{equation}
m(a_1) = 1 - \alpha\frac{\HH(\bxi_1)}{3}a_1\log(a_1) + \alpha a_1\left( \frac{4\pi}{3}\RR_s(\bxi_1;\bxi_1) + \frac{\HH(\bxi_1)}{3} \left(\frac{1}{3} -\log(\alpha)\right)\right) + O\left(a_1^2\right)\,.
\end{equation}
On the domain $\tilde{\Omega}$ with $O(1)$ length-scale this yields
\begin{equation}
m(a_1) = 1 - \frac{\tilde{H}(\tx_1)}{3}a_1\log(a_1) + a_1\left( \frac{4\pi}{3}\tilde{R}_s(\tx_1;\tx_1) + \frac{\tilde{H}(\tx_1)}{9}\right) + O\left(a_1^2\right)\,,
\end{equation}
and our final approximate integral solutions are
\begin{subequations}
\begin{align}
\int_{\Omega_{\bxi}} \GG_s(\bxi;\bxi_1)\PP(\bxi) d\bxi &\approx \PP(\bxi_1) \frac{\left(\alpha a_1\right)^2}{2}m(a_1)\,, \\
\int_{\Omega_{\bxi}} \GG_s(\bxi;\bxi_2)\PP(\bxi) d\bxi &\approx \PP(\bxi_1) \frac{2\pi}{3}a_1^3\alpha^3 \GG_s(\bxi_1;\bxi_2)\,.
\end{align}
\end{subequations}
%%%%%%%%%%%%%%%%%%%%%%%%%%%%%%%%%%%%%%%%%%%%%%%%%%%%%%%%%%%%%%%%%%%%%%%%%%%%%%%%%
\section{Reduced ionic and voltage asymptotic formulas}
%%%%%%%%%%%%%%%%%%%%%%%%%%%%%%%%%%%%%%%%%%%%%%%%%%%%%%%%%%%%%%%%%%%%%%%%%%%%%%%%%
%%%%%%%%%%%%%%%%%%%%%%%%%%%%%%%%%%%%%%%%%%%%%%%%%%%%%%%%%%%%%%%%%%%%%%%%%%%%%%%%%
\subsection{Small interior charge}\label{sec:small_alpha_app}
%%%%%%%%%%%%%%%%%%%%%%%%%%%%%%%%%%%%%%%%%%%%%%%%%%%%%%%%%%%%%%%%%%%%%%%%%%%%%%%%%
When $\alpha \sim O(1)$, we can expand the ratio $\AA$ assuming $a_1,\, a_2 \ll 1$ as
\begin{align*}
\AA &\approx
\frac{\left(\frac{f(a_1)}{a_1}-\pi\alpha\GG_s(\bxi_2;\bxi_1)\right)\left(1+\frac{g(a_2)}{4\alpha a_2}\left|\Omega_{\bxi}\right|\right) + \left(\frac{\pi g(a_2)}{4a_2} - \pi\alpha\GG_s(\bxi_2;\bxi_1)\right)\left(1 + \left|\Omega_{\bxi}\right|\GG_s(\bxi_2;\bxi_1)\right)}{\left(1+\frac{1}{2}\left(\alpha a_1\right)^2m(a_1) \right)\left(1+\frac{g(a_2)}{4\alpha a_2}\volOmxi\right)} \,, \\
&\approx \frac{f(a_1)}{a_1}-\pi\alpha\GG_s(\bxi_2;\bxi_1) + \frac{ \left(\frac{\pi g(a_2)}{4a_2} - \pi\alpha\GG_s(\bxi_2;\bxi_1)\right)\left(1 + \left|\Omega_{\bxi}\right|\GG_s(\bxi_2;\bxi_1)\right)}{\left(1+\frac{1}{2}\left(\alpha a_1\right)^2m(a_1) \right)\left(1+\frac{g(a_2)}{4\alpha a_2}\left|\Omega_{\bxi}\right|\right)} \,, \\
&\approx \frac{f(a_1)}{a_1}-\pi\alpha\GG_s(\bxi_2;\bxi_1) + \frac{ \left(\frac{\pi \alpha}{\volOmxi} - \frac{4a_2\pi\alpha\GG_s\left(\bxi_2;\bxi_1\right)}{g(a_2)\volOmxi}\right)\left(1 + \left|\Omega_{\bxi}\right|\GG_s(\bxi_2;\bxi_1)\right)}{1 + \frac{4\alpha a_2}{g(a_2)\left|\Omega_{\bxi}\right|}} \,, \\
&\approx \frac{f(a_1)}{a_1} - \pi\alpha\GG_s(\bxi_2;\bxi_1) + \frac{\pi \alpha}{\volOmxi} + \pi\alpha\GG_s(\bxi_2;\bxi_1) + O(a_2)\,, \\
&\approx \frac{f(a_1)}{a_1} + \frac{\pi \alpha}{\volOmxi} + O(a_2) \,,
\end{align*}
and similarly for $\BB$, which becomes
\begin{align*}
\BB &\approx
\frac{\frac{\pi g(a_2)}{4a_2} - \pi\alpha\GG_s(\bxi_2;\bxi_1)}
{1+\frac{g(a_2)}{4\alpha a_2}\volOmxi}
+
\frac{\left(\frac{f(a_1)}{a_1}-\pi\alpha\GG_s(\bxi_2;\bxi_1)\right)\frac{2\pi}{3}a_1^3\alpha^3 \GG_s(\bxi_1;\bxi_2)}
{\left(1+\frac{1}{2}\left(\alpha a_1\right)^2m(a_1) \right)\left(1+\frac{g(a_2)}{4\alpha a_2}\volOmxi\right)} \,, \\
&\approx \frac{\frac{\pi \alpha}{\volOmxi} - \frac{4a_2 \pi \alpha \GG_s\left(\bxi_2;\bxi_1\right)}{g(a_2)\volOmxi}}
{1+\frac{4\alpha a_2}{g(a_2)\volOmxi}}\,, \\
&\approx \frac{\pi \alpha}{\volOmxi} + O(a_2)\,.
\end{align*}
Starting from the general voltage formulas Eq.~\eqref{eq:VV_gen}, we obtain the current-voltage relation when $\alpha \sim O(1)$ by performing the asymptotic reduction below
\begin{align*}
\VV(\bxi_1) &= Ja_1^2\left( \left( m(a_1) - \frac{4\pi}{3}a_1\alpha\GGREENEVAL\right)\frac{1}{2}(\alpha a_1)^2\AA + \left(\frac{g(a_2)}{4 a_2} - \alpha\GG_s(\bxi_2;\bxi_1) \right) \frac{\volOmxi}{\alpha}\BB \right) + O\left(\left(\frac{J}{\alpha}\right)^2 \right)\,, \\
&\approx Ja_1^2\left(O(a_1) + \left(\frac{1}{4a_2} - \alpha\frac{\HH(\bxi_2)}{4\pi} \log(a_2) + O(1) \right)\pi\right) + O\left(\left(\frac{J}{\alpha}\right)^2 \right) \,, \\
&\approx Ja_1^2\left(\frac{\pi}{4a_2} - \alpha\frac{\HH(\bxi_2)}{4} \log(a_2) + O(1) \right) + O\left(\left(\frac{J}{\alpha}\right)^2 \right)\,,
\end{align*}
as well as, for the spatial average
\begin{align*}
\overline{\VV} &= Ja_1^2\left(\volOmxi\frac{g(a_2)}{4\alpha a_2}\BB - \frac{2\pi}{3}a_1^3\alpha^3\GGREENEVAL\AA\right) + O\left(\left(\frac{J}{\alpha}\right)^2 \right) \,, \\
&\approx Ja_1^2\left(\volOmxi\frac{g(a_2)}{4\alpha a_2} \left(\frac{\frac{\pi g(a_2)}{4 a_2} - \pi\alpha\GG_s(\bxi_2;\bxi_1) + O(a_1^3)}{1 + \frac{g(a_2)}{4\alpha a_2}\volOmxi + O(a_1^3)} \right) + O(a_1^2)\right) + O\left(\left(\frac{J}{\alpha}\right)^2 \right) \,, \\
&\approx Ja_1^2 \left( \frac{\frac{\pi g(a_2)}{4 a_2} - \pi\alpha\GG_s(\bxi_2;\bxi_1)}{1 + O(a_2)} + O(a_1^2) \right) + O\left(\left(\frac{J}{\alpha}\right)^2 \right) \,, \\
&\approx Ja_1^2 \left( \frac{\pi}{4a_2} - \alpha\frac{\HH(\bxi_2)}{4} \log(a_2) + O(1) \right) + O\left(\left(\frac{J}{\alpha}\right)^2 \right) \,,
\end{align*}
and thus we get the same expression for both $\VV(\x_1)$ and $\overline{\VV}$.
%%%%%%%%%%%%%%%%%%%%%%%%%%%%%%%%%%%%%%%%%%%%%%%%%%%%%%%%%%%%%%%%%%%%%%%%%%%%%%%%%
\subsection{Large interior charge}\label{sec:big_alpha_app}
%%%%%%%%%%%%%%%%%%%%%%%%%%%%%%%%%%%%%%%%%%%%%%%%%%%%%%%%%%%%%%%%%%%%%%%%%%%%%%%%%
Alternatively if $\alpha \gg 1$, we first remark that
\begin{equation}
\volOmxi = \alpha^3 \atOmega\,, \quad \GGREENEVAL = \alpha \ggreeneval\,,
\end{equation}
and thus we obtain that
\begin{align*}
\AA &=
\frac{1}{\alpha^2}\frac{\left(\frac{f(a_1)}{a_1}-\pi \tilde{G}_s(\tx_2;\tx_1)\right)\left(\frac{1}{\alpha^2} + \frac{g(a_2)}{4a_2}\atOmega\right) + \left(\frac{\pi g(a_2)}{4a_2} - \pi \tilde{G}_s(\tx_2;\tx_1)\right)\left(\frac{1}{\alpha^2} + \atOmega \tilde{G}_s\left(\tx_2;\tx_1\right)\right)}
{\left(\frac{1}{\alpha^2}+\frac{a_1^2}{2}m(a_1) \right)\left(\frac{1}{\alpha^2}+\frac{g(a_2)}{4a_2}\atOmega\right) - \left(\frac{1}{\alpha^2} + \atOmega \tilde{G}_s(\tx_2;\tx_1)\right)\frac{2\pi}{3}a_1^3 \tilde{G}_s(\tx_1;\tx_2)} \,.
\end{align*}
Taking first the limit of $\alpha$ big, we find
\begin{align*}
\AA &\approx \frac{1}{\alpha^2}\frac{\left(\frac{f(a_1)}{a_1}-\pi \tilde{G}_s(\tx_2;\tx_1)\right) \frac{g(a_2)}{4a_2} + \left(\frac{\pi g(a_2)}{4a_2} - \pi \tilde{G}_s(\tx_2;\tx_1)\right)\tilde{G}_s\left(\tx_2;\tx_1\right)}
{\frac{a_1^2}{8a_2}m(a_1)g(a_2) - \tilde{G}_s(\tx_2;\tx_1) \frac{2\pi}{3}a_1^3 \tilde{G}_s(\tx_1;\tx_2)} \,,
\end{align*}
after which we can expand assuming $a_1 \ll 1$ and $a_2 \ll 1$, leading to
\begin{align*}
\AA &\approx \frac{2}{\alpha^2 a_1^3}\left(\frac{f(a_1) - 4 \pi \frac{a_1 a_2}{g(a_2)} \left(\tilde{G}_s(\tx_2;\tx_1)\right)^2}
{m(a_1) - \frac{16\pi}{3}\frac{a_1a_2}{g(a_2)}\left(\tilde{G}_s(\tx_2;\tx_1)\right)^2}\right) \approx \frac{2}{\alpha^2 a_1^3}\left(1 + O(a_1\log(a_1)) \right) \,,
\end{align*}
and then if $a_1^3 \gg \frac{1}{\alpha^2}$ we can neglect the ratio $\AA$. We proceed similarly for $\BB$,
\begin{align*}
\BB &=
\frac{1}{\alpha^2}\frac{\left(\frac{1}{\alpha^2} + \frac{a_1^2}{2}m(a_1) \right)\left(\frac{\pi g(a_2)}{4a_2} - \pi \tilde{G}_s(\tx_2;\tx_1)\right) + \left(\frac{f(a_1)}{a_1}-\pi \tilde{G}_s(\tx_2;\tx_1)\right)\frac{2\pi}{3}a_1^3 \tilde{G}_s(\tx_1;\tx_2)}
{\left(\frac{1}{\alpha^2}+\frac{a_1^2}{2}m(a_1) \right)\left(\frac{1}{\alpha^2}+\frac{g(a_2)}{4a_2}\atOmega\right) - \left(\frac{1}{\alpha^2} + \atOmega \tilde{G}_s(\tx_2;\tx_1)\right)\frac{2\pi}{3}a_1^3 \tilde{G}_s(\tx_1;\tx_2)} \,,
\end{align*}
and then expanding in $\alpha \gg 1$ yields
\begin{align*}
\BB &\approx \frac{1}{\alpha^2\atOmega}\frac{\frac{a_1^2}{2}m(a_1) \left(\frac{\pi g(a_2)}{4a_2} - \pi \tilde{G}_s(\tx_2;\tx_1)\right) + \left(\frac{f(a_1)}{a_1}-\pi \tilde{G}_s(\tx_2;\tx_1)\right)\frac{2\pi}{3}a_1^3 \tilde{G}_s(\tx_1;\tx_2)}
{\frac{a_1^2}{8a_2}m(a_1)g(a_2) - \tilde{G}_s(\tx_2;\tx_1)\frac{2\pi}{3}a_1^3 \tilde{G}_s(\tx_1;\tx_2)} \,,
\end{align*}
and after some simplifications and keeping only leading-order terms in $a_1 \ll 1$ and $a_2 \ll 1$, we get
\begin{align*}
\BB &\approx
\frac{\pi}{\alpha^2 \atOmega} \left(\frac{g(a_2) - 4a_2\tilde{G}_s(\tx_2;\tx_1) + \frac{16a_2}{3m(a_1)}\left(f(a_1) - \pi a_1 \tilde{G}_s(\tx_2;\tx_1)\right) \tilde{G}_s(\tx_1;\tx_2)}
{g(a_2) - \frac{16\pi a_1a_2}{3m(a_1)}\left(\tilde{G}_s(\tx_2;\tx_1)\right)^2}\right)\,, \\
&\approx \frac{\pi}{\alpha^2 \atOmega}\left(1 + O(a_2\log(a_2)) \right) \,,
\end{align*}
since $\alpha \gg 1$. Concerning the voltage formula Eq.~\eqref{eq:VV_gen_influx}, we consider first the contribution from,
\begin{align*}
& \left( m(a_1) - \frac{4\pi}{3}a_1\ggreeneval\right)\frac{1}{2}(\alpha a_1)^2\AA\,,
\end{align*}
on which we can take the limit $\alpha \gg 1$, leading to
\begin{align*}
& \left( m(a_1) - \frac{4\pi}{3}a_1\ggreeneval\right)\left(
\frac{\left(\frac{f(a_1)}{a_1}-\pi \tilde{G}_s(\tx_2;\tx_1)\right)\frac{g(a_2)}{4a_2} + \left(\frac{\pi g(a_2)}{4a_2} - \pi \tilde{G}_s(\tx_2;\tx_1)\right) \tilde{G}_s\left(\tx_2;\tx_1\right)}
{m(a_1) \frac{g(a_2)}{4a_2} - \frac{4\pi}{3}a_1 \left(\tilde{G}_s(\tx_2;\tx_1)\right)^2}\right)\,,
\end{align*}
which is equivalent to
\begin{align*}
& \left( m(a_1) - \frac{4\pi}{3}a_1\ggreeneval\right)\left(
\frac{\left(\frac{f(a_1)}{a_1}\right)\frac{g(a_2)}{4a_2}  - \pi \left(\tilde{G}_s(\tx_2;\tx_1)\right)^2}
{m(a_1) \frac{g(a_2)}{4a_2} - \frac{4\pi}{3}a_1 \left(\tilde{G}_s(\tx_2;\tx_1)\right)^2}\right) \approx \\
& \left(1 - \frac{4\pi a_1}{3m(a_1)}\ggreeneval\right)\left(
\frac{\frac{f(a_1)}{a_1}  - \frac{4\pi a_2}{g(a_2)} \left(\tilde{G}_s(\tx_2;\tx_1)\right)^2}
{1 - \frac{16\pi a_1a_2}{3m(a_1)g(a_2)}\left(\tilde{G}_s(\tx_2;\tx_1)\right)^2}\right)
\,,
\end{align*}
and then neglecting terms of $O(1)$ and beyond in terms of the narrow radii $a_1$ and $a_2$ we get,
\begin{align*}
\left( m(a_1) - \frac{4\pi}{3}a_1\ggreeneval\right)\frac{1}{2}(\alpha a_1)^2\AA \approx \frac{1}{a_1} - \frac{\tilde{H}(\tx_1)}{4}\log(a_1) + O(1)\,.
\end{align*}
Similarly for the contribution from
\begin{align*}
\left(\frac{g(a_2)}{4 a_2} - \ggreeneval \right) \alpha^2 \atOmega \BB\,,
\end{align*}
we get after taking the limit $\alpha \gg 1$,
\begin{align*}
\left(\frac{g(a_2)}{4 a_2} - \ggreeneval \right) \left( \frac{\frac{a_1^2}{2}m(a_1)\left(\frac{\pi g(a_2)}{4a_2} - \pi \tilde{G}_s(\tx_2;\tx_1)\right) + \left(\frac{f(a_1)}{a_1}-\pi \tilde{G}_s(\tx_2;\tx_1)\right)\frac{2\pi}{3}a_1^3 \tilde{G}_s(\tx_1;\tx_2)}
{\frac{a_1^2}{2}m(a_1)\frac{g(a_2)}{4a_2} - \frac{2\pi}{3}a_1^3 \left(\tilde{G}_s(\tx_2;\tx_1)\right)^2}\right) \,,
\end{align*}
and then further simplifications yield
\begin{align*}
\left(g(a_2) - 4a_2\ggreeneval \right) \left( \frac{\frac{\pi g(a_2)}{4a_2} - \pi \tilde{G}_s(\tx_2;\tx_1) + \left(\frac{f(a_1)}{a_1}-\pi \tilde{G}_s(\tx_2;\tx_1)\right)\frac{4\pi a_1}{3m(a_1)} \tilde{G}_s(\tx_1;\tx_2)}
{g(a_2) - \frac{16\pi a_1a_2}{3m(a_1)} \left(\tilde{G}_s(\tx_2;\tx_1)\right)^2}\right) \,,
\end{align*}
and upon neglecting terms of order $O(1)$ and beyond we get
\begin{align*}
\left(\frac{g(a_2)}{4 a_2} - \ggreeneval \right) \alpha^2 \atOmega \BB \approx \frac{\pi}{4a_2} - \frac{\tilde{H}(\tx_2)}{4}\log(a_2) + O(1)\,.
\end{align*}
Upon adding the two contributions we obtain
\begin{equation}
\tv(\tx_1) \approx Ja_1^2 \left(\frac{1}{a_1} + \frac{\pi}{4a_2} - \frac{\tilde{H}(\tx_1)}{4}\log(a_1) - \frac{\tilde{H}(\tx_2)}{4}\log(a_2) + O(1) \right)\,.
\end{equation}
Similarly for the average voltage with formula given by Eq.~\eqref{eq:VV_gen_bar}, we first calculate
\begin{align*}
\frac{g(a_2)}{4a_2}\alpha^2\atOmega \BB \approx \frac{g(a_2)}{4a_2}\left(\frac{\frac{a_1^2}{2}m(a_1) \left(\frac{\pi g(a_2)}{4a_2} - \pi \tilde{G}_s(\tx_2;\tx_1)\right) + \left(\frac{f(a_1)}{a_1}-\pi \tilde{G}_s(\tx_2;\tx_1)\right)\frac{2\pi}{3}a_1^3 \tilde{G}_s(\tx_1;\tx_2)}
{\frac{a_1^2m(a_1)}{2}\frac{g(a_2)}{4a_2} - \tilde{G}_s(\tx_2;\tx_1)\frac{2\pi}{3}a_1^3 \tilde{G}_s(\tx_1;\tx_2)}\right)\,,
\end{align*}
and then further simplifications yield
\begin{align*}
\frac{g(a_2)}{4a_2}\alpha^2\atOmega \BB \approx \frac{\frac{\pi g(a_2)}{4a_2} - \pi \tilde{G}_s(\tx_2;\tx_1) + \left(\frac{f(a_1)}{a_1}-\pi \tilde{G}_s(\tx_2;\tx_1)\right)\frac{4\pi a_1}{3m(a_1)}\tilde{G}_s(\tx_1;\tx_2)}{1 + \frac{16\pi a_1a_2}{3m(a_1)g(a_2)}\left(\ggreeneval\right)^2}
\end{align*}
and then neglecting terms of order $O(1)$ and beyond, we are left with
\begin{align*}
\frac{g(a_2)}{4a_2}\alpha^2\atOmega \BB \approx \frac{\pi}{4a_2} - \frac{\tilde{H}(\tx_2)}{4}\log(a_2) + O(1)\,.
\end{align*}
Similarly for the term
\begin{align*}
-\frac{2\pi}{3}a_1^3\alpha^2 \tilde{G}_s(\tx_1;\tx_2)\AA &\approx -\frac{2\pi}{3}a_1^3\tilde{G}_s(\tx_1;\tx_2) \left(\frac{\left(\frac{f(a_1)}{a_1}-\pi \tilde{G}_s(\tx_2;\tx_1)\right) \frac{g(a_2)}{4a_2} + \left(\frac{\pi g(a_2)}{4a_2} - \pi \tilde{G}_s(\tx_2;\tx_1)\right)\tilde{G}_s\left(\tx_2;\tx_1\right)}
{\frac{a_1^2}{2}m(a_1)\frac{g(a_2)}{4a_2} - \tilde{G}_s(\tx_2;\tx_1) \frac{2\pi}{3}a_1^3 \tilde{G}_s(\tx_1;\tx_2)}\right) \,, \\
&\approx -\frac{4\pi}{3}a_1\tilde{G}_s(\tx_1;\tx_2) \left(\frac{\frac{f(a_1)}{a_1} - \frac{4\pi a_2}{g(a_2)} \left(\tilde{G}_s(\tx_2;\tx_1)\right)^2}
{m(a_1) - \frac{16\pi a_1a_2}{3g(a_2)} \left(\tilde{G}_s(\tx_1;\tx_2)\right)^2}\right)\,, \\
&\approx - \frac{4\pi}{3}\tilde{G}_s(\tx_1;\tx_2) \sim O(1) \,.
\end{align*}
Thus for the average voltage we are left with
\begin{equation}
\overline{\tv} = Ja_1^2 \left( \frac{\pi}{4a_2} - \frac{\tilde{H}(\tx_2)}{4}\log(a_2) + O(1) \right)\,.
\end{equation}
%%%%%%%%%%%%%%%%%%%%%%%%%%%%%%%%%%%%%%%%%%%%%%%%%%%%%%%%%%%%%%%%%%%%%%%%%%%%%%%%%
\section{Numerical solutions of electro-diffusion model using COMSOL Multiphysics}\label{sec:com}
\begin{figure}[!ht]
\centering
\includegraphics[width=\textwidth]{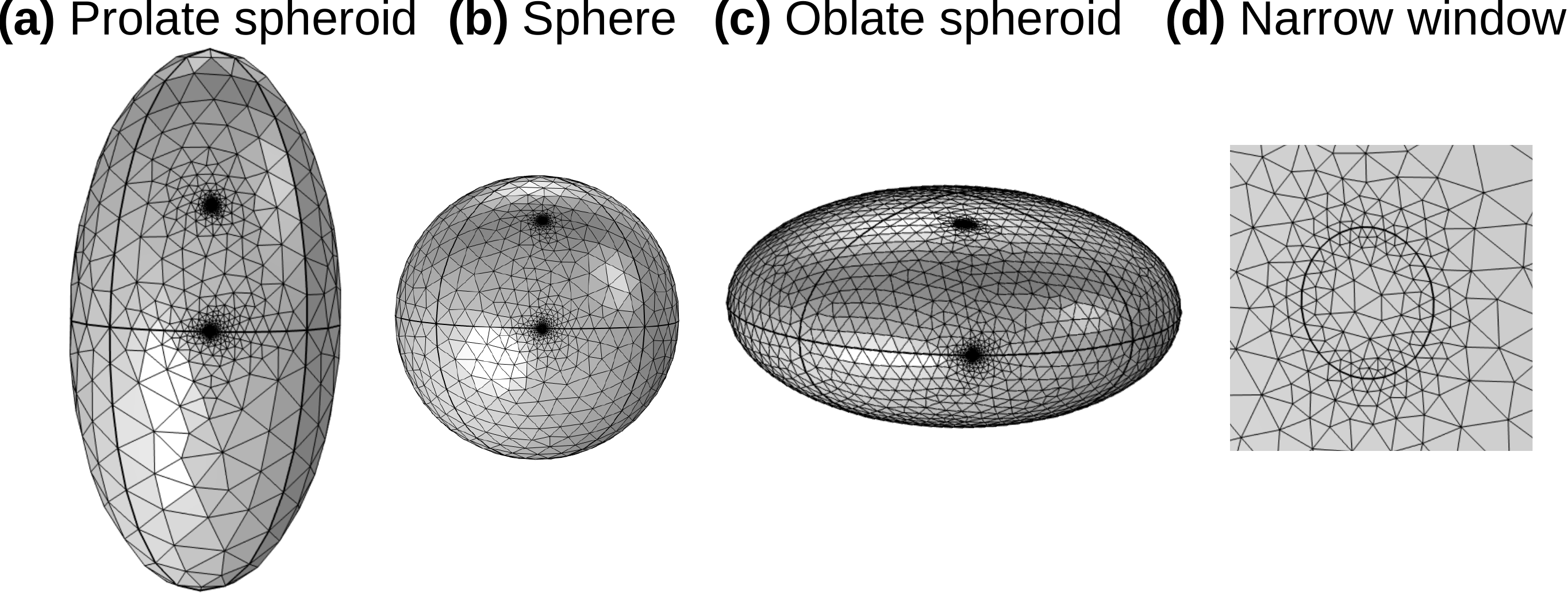}
\caption{\label{fig:mesh} \textbf{Spheroid domain mesh obtained with COMSOL \cite{comsol}.} \textbf{(a)} Prolate spheroid. \textbf{(b)} Sphere. \textbf{(c)} Oblate spheroid. \textbf{(d)} Zoom on the refined mesh of a narrow window.}
\end{figure}

%%%%%%%%%%%%%%%%%%%%%%%%%%%%%%%%%%%%%%%%%%%%%%%%%%%%%%%%%%%%%%%%%%%%%%%%%%%%%%%%%
We numerically solve the steady-state system of Poisson-Nernst-Planck Eq.~\eqref{eq:ss_eqs} with mixed boundary conditions Eq.~\eqref{eq:pnp_BC_tilde} using the \textit{Coefficient Form PDE} module from COMSOL Multiphysics version 6.1 \cite{comsol}. The spheroid domains are discretized with a free tetrahedral mesh that is \textit{extremely fine} on the narrow windows and \textit{fine} for the rest of the domain (Fig.~\ref{fig:mesh}). The predefined mesh preferences from COMSOL, which control the maximum and minimum element sizes, vary from \textit{extremely coarse} to \textit{extremely fine}.
%%%%%%%%%%%%%%%%%%%%%%%%%%%%%%%%%%%%%%%%%%%%%%%%%%%%%%%%%%%%%%%%%%%%%%%%%%%%%%%%%
% Perhaps redo figure with a normalization using the long axis to be consistent.
%%%%%%%%%%%%%%%%%%%%%%%%%%%%%%%%%%%%%%%%%%%%%%%%%%%%%%%%%%%%%%%%%%%%%%%%%%%%%%%%%
\end{appendix}
%%%%%%%%%%%%%%%%%%%%%%%%%%%%%%%%%%%%%%%%%%%%%%%%%%%%%%%%%%%%%%%%%%%%%%%%%%%%%%%%%
%%%%%%%%%%%%%%%%%%%%%%%%%%%%%%%%%%%%%%%%%%%%%%%%%%%%%%%%%%%%%%%%%%%%%%%%%%%%%%%%%
\normalem
% \bibliographystyle{ieeetr}
% \bibliography{ref}

%%%%%%%%%%%%%%%%%%%%%%%%%%%%%%%%%%%%%%%%%%%%%%%%%%%%%%%%%%%%%%%%%%%%%%%%%%%%%%%%%
\end{document}